\documentclass[11pt]{article}

\usepackage[utf8]{inputenc}
\usepackage[T1]{fontenc}
\usepackage{graphicx}
\usepackage{grffile}
\usepackage{longtable}
\usepackage{wrapfig}
\usepackage[figuresright]{rotating}
\usepackage[normalem]{ulem}
\usepackage{amsmath}
\usepackage{textcomp}
\usepackage{amssymb}
\usepackage{capt-of}
\usepackage{hyperref}
\usepackage{arxiv}
\usepackage{newtxtext,newtxmath}
\makeatletter
\renewcommand\normalsize{%
\@setfontsize\normalsize\@xipt{13.6}%
\abovedisplayskip 11\p@ \@plus3\p@ \@minus6\p@
\abovedisplayshortskip \z@ \@plus3\p@
\belowdisplayshortskip 6.5\p@ \@plus3.5\p@ \@minus3\p@
\belowdisplayskip \abovedisplayskip}
\normalsize
\renewcommand\small{%
\@setfontsize\small\@xpt\@xiipt
\abovedisplayskip 10\p@ \@plus2\p@ \@minus5\p@
\abovedisplayshortskip \z@ \@plus3\p@
\belowdisplayshortskip 6\p@ \@plus3\p@ \@minus3\p@
\belowdisplayskip \abovedisplayskip}
\renewcommand\footnotesize{\@setfontsize\footnotesize\@ixpt{11}}
\renewcommand\scriptsize{\@setfontsize\scriptsize\@viiipt{9.5}}
\renewcommand\tiny{\@setfontsize\tiny\@vipt\@viipt}
\renewcommand\large{\@setfontsize\large\@xiipt{14.5}}
\renewcommand\Large{\@setfontsize\Large\@xivpt{18}}
\renewcommand\LARGE{\@setfontsize\LARGE\@xviipt{22}}
\renewcommand\huge{\@setfontsize\huge\@xxpt{25}}
\renewcommand\Huge{\@setfontsize\Huge\@xxvpt{30}}
\makeatother
\usepackage{textcomp}
\usepackage{authblk}

\usepackage{booktabs}
\usepackage{threeparttable}
\usepackage{tabularx}
\usepackage{threeparttablex}
\usepackage{siunitx}
\usepackage{makecell}
\usepackage{multirow}
\usepackage{placeins}
\usepackage{float}
\usepackage{pdflscape}
\usepackage{upquote}
\usepackage[american, english]{babel}
\usepackage{csquotes}
\usepackage[sortcites=true,sorting=nyt,backend=biber,annotation=false,style=apa]{biblatex}

\DeclareLanguageMapping{american}{american-apa}
\AtBeginBibliography{\small}
\definecolor{navyblue}{rgb}{0.0, 0.0, 0.5}
\usepackage{hyperref}
\hypersetup{colorlinks=true, linkcolor=black, citecolor=navyblue, urlcolor=navyblue}
\usepackage{listings}
\author[1,2]{Dominik Pegler\thanks{Corresponding author: dominik.pegler@univie.ac.at}}
\author[2]{Frank Jäkel}
\author[1]{David Steyrl}
\author[1]{Frank Scharnowski}
\author[1]{Filip Melinscak}
\affil[1]{Department of Cognition, Emotion, and Methods in Psychology, Faculty of Psychology, University of Vienna}
\affil[2]{Centre for Cognitive Science \& Institute of Psychology, TU Darmstadt}
\renewcommand{\shorttitle}{Unpacking Interpretability}
\date{}
\title{Unpacking Interpretability: Human-Centered Criteria for Optimal Combinatorial Solutions}
\hypersetup{
 pdfauthor={},
 pdftitle={Unpacking Interpretability: Human-Centered Criteria for Optimal Combinatorial Solutions},
 pdfkeywords={},
 pdfsubject={},
 pdfcreator={},
 pdflang={English}}
\begin{document}

\maketitle
\begin{abstract}
Algorithmic support systems can return solutions that are optimal but hard to interpret for humans. A given problem can have multiple, equally optimal solutions that differ substantially in how interpretable they are. Yet there is no precise account of what makes one solution more interpretable than another. To identify structural properties associated with interpretability, we studied a packing paradigm based on the multiple subset sum problem (MSSP), in which participants judged which of two equally optimal solutions for packing items into bins was easier to understand. We found that preferences reliably tracked three properties of solution structure: alignment with a greedy heuristic, simple within-bin composition, and how ordered the visual representation was. The strongest associations were observed for ordered representations and alignment with a greedy heuristic. Compositional simplicity also showed a consistent association. Reaction-time evidence was mixed; the only reliable result was faster responses when heuristic differences were larger. Webcam-based results of aggregate gaze times did not show reliable associations with the three solution properties. In summary, these results ground interpretability (at least in our packing paradigm) in measurable structural properties of solutions. By identifying these properties (simple compositions, ordered representation, and alignment with a heuristic) our findings motivate interpretability-aware optimization and presentation of machine solutions. They also outline a path to quantify trade-offs between optimality and interpretability in allocation and design tasks.
\end{abstract}


\addtocontents{toc}{\protect\setcounter{tocdepth}{-1}}
\begin{refsection}
\section{Introduction}
\label{sec:org8d31d2a}

Advances in algorithmic optimization and machine learning increasingly place automated solvers at the center of human–machine collaboration \autocite{akataResearchAgendaHybrid2020,krakowskiArtificialIntelligenceChanging2023}. In many real deployments, when these solvers produce plans or assignments, human interpretability becomes a practical prerequisite for adoption and safe use. Many optimization problems admit multiple solutions that are equally optimal but differ substantially in their structure and presentation. The open research question is therefore: when optimal solutions are tied on value, which structural properties make one solution easier to understand than another? To isolate such properties under controlled conditions, we study this question in a deliberately simple but nontrivial packing paradigm in which multiple distinct solutions can be equally optimal yet differ markedly in how understandable they seem to people.
\subsection{Combinatorial packing and the multiple subset sum problem (MSSP)}
\label{sec:org1674241}

\begin{figure}[htbp]
  \centering \includegraphics[height=.35\textheight]{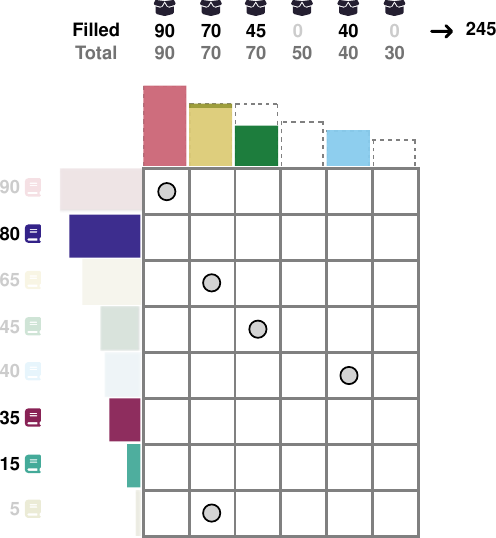}
  \caption{\label{fig:problem}An instance of the Multiple Subset Sum Problem (MSSP). Rows denote items, and columns denote bins. Item assignments are indicated by gray dots in cells. Item sizes are represented by block lengths and numerical labels. The overall objective score (total packed size) is shown in the upper-right corner.}\end{figure}

Packing problems --- such as the classical bin packing problem \autocite{johnsonWorstcasePerformanceBounds1974} and multi-knapsack \autocite{cacchianiKnapsackProblemsOverview2022} --- require assigning items of varying sizes to capacity-limited bins under hard constraints. This class is foundational in operations research and has high-impact applications in resource allocation and logistics \autocite{gunawanTrendsMultidisciplinaryScheduling2021}. For example, hospitals have to assign patients (items with care requirements) to a limited number of nurses (bins with capacities) \autocite{marzoukNursePatientAssignment2021}. Capital budgeting similarly requires allocating limited resources across competing projects \autocite{gurskiKnapsackProblemsParameterized2019}. We study the multiple subset sum problem (MSSP; \cite{capraraMultipleSubsetSum2000}), a special case of multi-knapsack in which each item's profit equals its size, the number and capacity of bins are fixed, and the objective is to maximize total packed size. We chose MSSP because it preserves the central feature needed for this study --- multiple structurally different solutions with identical objective value --- while keeping the task controlled enough to compare interpretable structure without confounding it with objective quality. Figure \ref{fig:problem} illustrates an instance of the MSSP used in this study, visually represented as an assignment matrix.

\subsection{Interpreting optimal solutions}
\label{sec:org5fe5e9e}

Even when clearly presented, optimal solutions to combinatorial problems like the MSSP can vary substantially in how readily humans can grasp their underlying structure and rationale. Following established usage, we refer to this human-centered quality as interpretability: the degree to which users can understand and effectively work with a machine-generated solution \autocite{doshi-velezRigorousScienceInterpretable2017}. Psychologically, interpretability interacts with perception, understanding, and trust: people favor structures they can mentally simulate or justify, even at the cost of forgoing opaque but optimal alternatives \autocites{kahnemanThinkingFastSlow2011}[][]{millerExplanationArtificialIntelligence2019}[][]{liptonMythosModelInterpretability2017}[][]{tverskyJudgmentUncertaintyHeuristics1974}[][]{dietvorstAlgorithmAversionPeople2015}[][]{swellerCognitiveLoadProblem1988}[][]{zerilliHowTransparencyModulates2022}[][]{bussoneRoleExplanationsTrust2015}[][]{leeTrustAutomationDesigning2004}. While a substantial portion of explainable artificial intelligence (XAI) research has focused on explanations for predictions \autocites{barredoarrietaExplainableArtificialIntelligence2020}[][]{abdulTrendsTrajectoriesExplainable2018}[][]{rudinStopExplainingBlack2019}, far less is known about what makes one optimal solution more intelligible than another in combinatorial settings (but see \cite{ibsHumanExplanationsExplainable2024,ottSimplifExSimplifyingExplaining2023,ibsGeneratingRationalesBased2026}). Crucially, research on explanation interpretability has shown that increasing the complexity of explanations (e.g., through more terms or new concepts) can increase the time required for humans to verify their consistency \autocite{narayananHowHumansUnderstand2018}.
\subsection{Complexity-informed proxies for interpretability}
\label{sec:org670116d}

We focus on three solution-level properties that contribute to interpretability and align with well-established cognitive and perceptual principles. First, humans often rely on simple heuristics to solve problems, preferring structures that match familiar construction rules and finding large deviations harder to rationalize \autocites{gigerenzerHeuristicDecisionMaking2011}[][]{tverskyJudgmentUncertaintyHeuristics1974}[][]{cormenIntroductionAlgorithms2009}. Second, compositional simplicity reduces cognitive load: bins that are nearly empty or nearly full and contain few items are easier to encode and compare than bins with many items, or bins that are half full \autocites{swellerCognitiveLoadProblem1988}[][]{tverskyAnimationCanIt2002}. Third, perceptual organization favors ordered layouts; sequences that can be summarized by short rules (e.g., ``largest first'') are preferred under both the simplicity principle and the principle of empirical likelihood \autocites{feldmanSimplicityPrinciplePerception2016}[][]{vanderhelmSimplicityLikelihoodVisual2000}{chaterReconcilingSimplicityLikelihood1996}{helmholtzTreatisePhysiologicalOptics1962}.

\begin{figure*}[htbp]
  \centering \includegraphics[width=1.0\textwidth]{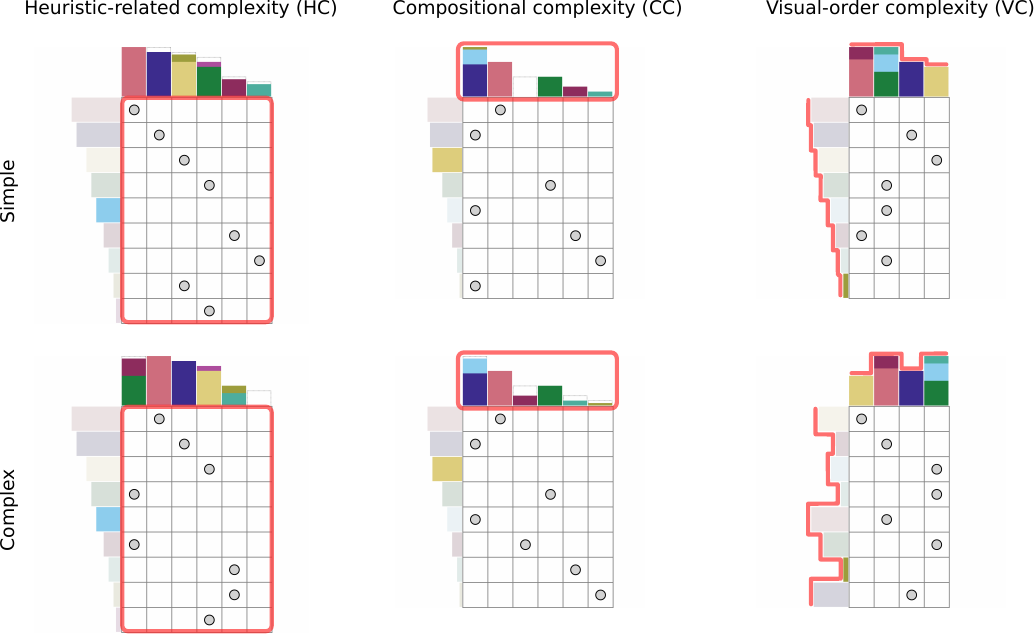}
  \caption{\label{fig:complexity}The three panels describe our intuition behind the three hypothesized complexity metrics. Each focuses on a different aspect of the solution, as highlighted by the red annotations. HC focuses on the assignments and how much they deviate from the greedy heuristic. CC focuses on the bins and how clean/organized/filled (see definition) they are. VC focuses on whether the elements of the solution are sorted by size.}\end{figure*}

As shown in Figure \ref{fig:complexity}, we operationalize these properties with three solution-level metrics, introduced here at a high level and defined in detail in \hyperref[methods]{Methods}. We selected these metrics because each maps onto a distinct, theory-grounded aspect of interpretability and each is directly actionable for optimization systems or interfaces. \emph{Heuristic-related complexity} (HC) quantifies deviation from a greedy packing heuristic, providing a measure of how closely a solution follows an intuitive construction. \emph{Compositional complexity} (CC) is intended to quantify how challenging a bin's contents are to grasp at a glance by combining information about the number of items in a bin, the balance of their sizes, and the amount of unused capacity, such that bins with many items and intermediate fill levels can in principle be treated as more complex than bins that are dominated by few items and are nearly empty or nearly full. \emph{Visual-order complexity} (VC) indexes the disorder of the display of bins and items, reflecting the degree to which a solution deviates from a sorted, rule-like presentation. Our metrics draw in part on information-theoretic principles --- CC, for instance, is defined as surprisal under a generative model --- but they are designed to capture specific cognitive mechanisms rather than generic statistical complexity. We do not claim that HC, CC, and VC exhaust all relevant determinants of interpretability; rather, they provide a theory-driven and operational starting set. As a visual-layout control, we include \emph{diagonal dissimilarity} (DD), a covariate that captures purely geometric similarity to an idealized diagonal-like assignment pattern.

\subsection{Prior work and gap}
\label{sec:orga23fe7d}

Explainable planning emphasizes aligning solutions with users' mental models --- through model reconciliation, contrastive rationales, or solution annotations --- highlighting that intelligibility depends on both derivation and presentation \autocites{foxExplainablePlanning2017}[][]{chakrabortiPlanExplanationsModel2017}. In bin packing and knapsack research, structural regularities and greedy heuristics are well-characterized \autocites{coffmanApproximationAlgorithmsBin1997}[][]{kellererKnapsackProblems2004}. Behavioral work shows that humans rely on simple strategies and that performance depends on instance structure \autocites{macgregorHumanPerformanceTraveling2011}[][]{dumnicPathGameCrowdsourcingTimeconstrained2019}[][]{murawskiHowHumansSolve2016}[][]{francoGenericPropertiesComputational2021}[][]{francoTaskindependentMetricsComputational2022}{ibsHumanExplanationsExplainable2024}. Even when asked to discriminate between solutions of varying optimality, humans may struggle to consistently identify the truly optimal option, suggesting inherent difficulties in evaluating complex combinatorial outputs \autocite{kyritsisPerceivedOptimalityCompeting2022}.

While explainable planning offers methods for justifying and aligning plans with users' mental models, and work on predictive explanations has advanced substantially in the field of XAI, empirical and feature-based accounts of interpretability for combinatorial optimization solutions are still emerging \autocite{ibsHumanExplanationsExplainable2024}. In parallel, work on cumulative cultural evolution in continuous optimization tasks shows that people come to prefer and reproduce solutions that match their inductive biases --- prior expectations or preferences for how a ``good'' solution should look, such as simplicity or symmetry --- and that misalignment between these biases and the true optimum can systematically limit collective performance \autocite{thompsonHumanBiasesLimit2021}. Related work in other optimization domains likewise examines preferences between competing solutions without specifying solution-level structural metrics \autocite{kyritsisPerceivedOptimalityCompeting2022}. Our study addresses this gap within the packing paradigm by quantifying how three solution-level properties --- HC, CC, and VC --- predict human choices, response speed, and attention when comparing equally optimal solutions.

Beyond describing features that shape interpretability, our practical aim is to enable interpretability-aware optimization. Embedding complexity metrics as secondary criteria --- e.g., tie-breaking among equal-value optima or soft penalties in multi-objective formulations --- would let optimizers return solutions that are both high in value and easy to understand \autocite{ehrgottMulticriteriaOptimization2005}.
\subsection{Pre-registered design, hypotheses and analysis plan}
\label{sec:org7836477}

\begin{figure*}[htbp]
  \centering \includegraphics[width=1.0\textwidth]{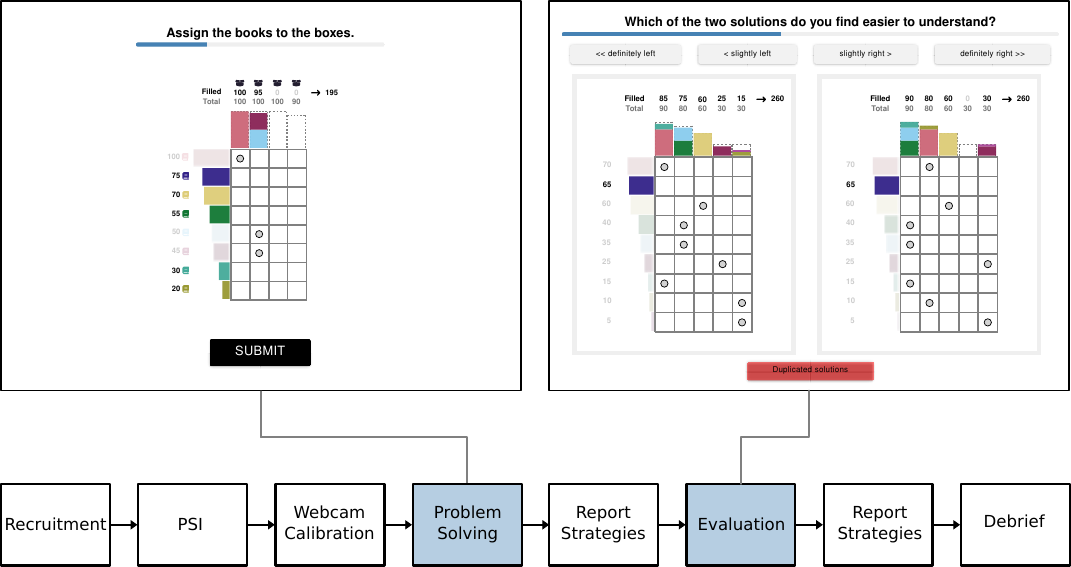}
  \caption{\label{fig:experiment}The diagram shows the study workflow including questionnaire (PSI = problem-solving inventory; \cite{heppnerDevelopmentImplicationsPersonal1982}), webcam calibration, seven problem-solving trials with feedback, and twenty-five evaluation trials. The left screen displays an example problem-solving trial. In evaluation trials (right screen), participants judged which of two optimal solutions was easier to understand ("definitely/slightly" left or right).}
\end{figure*}

We studied interpretability using the multiple subset sum problem introduced above \autocite{capraraMultipleSubsetSum2000,johnsonWorstcasePerformanceBounds1974}. Each instance contained several bins and items and was constructed to admit at least two distinct optimal solutions. Participants first practiced solving the task and received feedback. In the main evaluation phase, they viewed two equally optimal solutions to the same problem instance side by side and answered ``Which solution is easier to understand?'' on a four-level scale (definitely/slightly left/right), providing a direct behavioral measure of interpretability preference. Figure \ref{fig:experiment} summarizes the workflow.

We tested our three solution-level properties as drivers of interpretability preferences: HC, CC, and VC, with DD as a visual-layout control (full definitions in \hyperref[methods]{Methods}). Because participants evaluated pairs of equally optimal solutions, our analyses use between-solution differences derived from these solution-level metrics: signed right–left differences as predictors of choice and gaze, and absolute differences for reaction times. We complement preferences with two process measures: reaction times, which reflect overall processing effort and decisional conflict \autocite{luceResponseTimesTheir1986}, and webcam-based gaze. Webcam-based eye tracking provides aggregate dwell measures that indicate relative attention to the left versus right solution \autocites{papoutsakiWebgazerScalableWebcam2016}[][]{ecksteinEyeGazeWhat2017}[][]{gollanGazeQuantifyingConscious2025}. This framing allows us to link interpretable stimulus structure directly to behavioral preferences, processing speed, and attention.

We conducted an exploratory study to refine metrics followed by a preregistered confirmatory study using the fixed metrics. We tested three preregistered hypotheses. H1: within a pair of equally optimal solutions, participants prefer the option with lower HC, CC, and VC. H2: larger absolute between-solution differences in these metrics are associated with faster decisions. H3: more complex solutions attract relatively greater dwell time. While a general preference for simpler solutions may seem intuitive, it is not clear a priori which structural properties drive interpretability, whether they contribute independently, or how large their effects are. Quantifying these relationships is a prerequisite for systematically selecting or designing solutions that are both optimal and easy to understand. In the \hyperref[results]{Results} section, we report findings from the confirmatory sample and compare them with those from the exploratory sample.
\section{Methods}
\label{methods}
\subsection{The multiple subset sum problem (MSSP)}
\label{problem-definition-a-bin-packing-variant-multiple-subset-sum}
As introduced, our study focused on the multiple subset sum problem (MSSP; \cite{capraraMultipleSubsetSum2000}), a variant of the multi-knapsack problem \autocite{cacchianiKnapsackProblemsOverview2022} in which each item's profit equals its size. Given \(m\) bins with capacities \(w_{i}\) and \(n\) items with sizes \(z_{j}\), the task is to select a subset of items and assign each to at most one bin so as not to exceed any bin's capacity and to maximize the total packed size, as formulated in Equation \ref{eqn-obj}:

\begin{equation}\label{eqn-obj}
\begin{aligned}
\underset{x_{ij} \in \{0,1\}}{\text{maximize}} \quad &\sum_{i=1}^{m} \sum_{j=1}^{n} z_j x_{ij} , \\
\text{subject to} \quad &\sum_{j=1}^{n} z_{j}x_{ij} \leq w_{i}, \quad &&i = 1, 2, \ldots, m , \\
&\sum_{i=1}^{m} x_{ij} \leq 1, \quad &&j = 1, 2, \ldots, n .
\end{aligned}
\end{equation}

\noindent Here \(x_{ij}\) equals 1 if item \(j\) is placed in bin \(i\), and 0 otherwise. This fixed-bin, maximization objective differs from the classical bin packing objective of minimizing the number of bins. Figure \ref{fig:problem} illustrates an instance of our bin-packing variant and its experimental representation. Rows correspond to items and columns to bins. Block lengths and labels indicate item sizes \(z_{j}\), and filled cells (dots) in the assignment matrix indicate assignments \(x_{ij} = 1\). The number shown at the top right is the current score, i.e., the objective value \(\sum_{i,j}\ z_{j}\ x_{ij}\) in Equation \ref{eqn-obj}.

For the experiment, we randomly generated problem instances subject to several constraints. Instances contained 4--6 bins and 7--9 items, no item exceeded the largest bin, no bin was smaller than the smallest item, the ratio of total item size to total bin capacity lay between 0.8 and 1.0, and each instance admitted at least two distinct optimal solutions. These constraints were chosen to keep the task simple yet nontrivial. In particular, having more items than bins reduces the likelihood that each item maps to its own bin: an assignment that would be trivially interpretable and uninteresting to evaluate. This asymmetry also reflects many real-world allocation settings in which items outnumber bins \autocites{kellererKnapsackProblems2004}[][]{cacchianiKnapsackProblemsOverview2022}. Optimal solutions were generated with the OR-Tools CP-SAT solver (\texttt{ortools} 9.8.3296; \cite{cpsatlp}), using default solver settings, by iteratively excluding previously found optima until up to 100 distinct optimal solutions had been enumerated for an instance. A detailed description of stimulus generation is provided in the Supplementary Material (see ``Generation of problem and solution instances'').
\subsection{Overview of experimental design}
\label{overview-of-experimental-design}
Our web-based within-subjects design comprised two studies: an exploratory study used to refine the metrics and a preregistered confirmatory study used to test them. The studies were conducted in accordance with the Declaration of Helsinki and approved by the Ethics Committee of the University of Vienna (approval no. 01073, dated 27 February 2024); all participants provided electronic informed consent before participation. The full ethics approval information, including the coverage of both studies and the data-collection dates, is provided in the Ethical Statement at the end of the manuscript.

After providing informed consent, participants completed the Problem-Solving Inventory (PSI; \cite{heppnerDevelopmentImplicationsPersonal1982}) and received detailed instructions, including an interactive example of the task (see Fig. \ref{fig:experiment}). They then completed seven problem-solving trials with feedback, reported their problem-solving strategies in a free-text box, completed four practice evaluation trials and 25 actual evaluation trials, and finally reported their evaluation strategies and demographic information. 
\subsection{Experimental procedure}
\label{experimental-procedure}
\subsubsection{Participants}
\label{recruitment-of-participants}
The study was advertised on Prolific.com \autocite{palanProlificacASubjectPool2018} to participants residing in the US or UK who reported being fluent in English, having normal or corrected-to-normal vision, and possessing a webcam or built-in camera. Participants were compensated £9.00 per hour.

Following the exploratory study (see ``Exploratory study'' in the Supplementary Material), which involved 73 participants and 1,664 observations (evaluation trials), and uncovered a significant link between complexity and interpretability preferences, we used the same sample size for our confirmatory study. A total of 87 participants recruited from Prolific completed the confirmatory study, with 73 remaining after exclusion (exclusion rate = 16.1\%). Per-criterion exclusion counts for both studies are reported in the Supplementary Material (``Participant exclusions''). Ages ranged from 20 to 78 years (M = 45.00, SD = 12.65), and the sample consisted of 60.27\% male and 39.73\% female participants. Participants took a median of 24.72 minutes to complete the experiment (25th--75th percentile = 19.43--31.65 minutes). None of the participants in the exploratory study were permitted to participate in this confirmatory study.

For the gaze analyses, only participants with usable webcam-based eye-tracking data were included, resulting in 70 participants and 1,600 evaluation trials (see \hyperref[gaze-dwell-time-no-effects-of-complexity]{Gaze Dwell Times}) in the confirmatory study. Participants without any valid gaze samples, for example due to calibration or tracking failures, contributed only to the behavioral analyses.
\subsubsection{Eye-tracking calibration}
\label{eye-tracking-calibration}
Webcam-based eye tracking used the open-source JavaScript library WebGazer.js \autocite{papoutsakiWebgazerScalableWebcam2016}. Participants calibrated by fixating instructed screen points and received accuracy feedback; poor calibrations could be repeated before the experimental trials began.
\subsubsection{Experimental trials}
\label{experimental-trials}
\paragraph{Problem-solving trials}
\label{sec:orgc0ca0f5}
To become familiar with the task, participants first solved seven problem instances themselves (see Figure \ref{fig:experiment}). The instances were the same for all participants, increased in difficulty, and were followed by feedback indicating whether the submitted solution was optimal; if not, an optimal solution was shown side by side with the participant's own solution.
\paragraph{Evaluation trials}
\label{evaluation-trials}
In each of the 25 evaluation trials, participants viewed a pair of equally optimal solutions to the same problem and answered ``Which of the two solutions do you find easier to understand?'' using four response options: \emph{definitely left}, \emph{slightly left}, \emph{slightly right}, and \emph{definitely right} (see Figure \ref{fig:experiment}). There was no time limit. Two catch trials required participants to detect identical solutions using a fifth ``Duplicated solutions'' button, and three coherence trials assessed whether judgments across linked pairs were transitive. The catch and coherence trials were fixed across participants; the remaining 20 evaluation trials were sampled from a larger pool of possible pairs. A detailed description of trial generation and predictor distributions is provided in the Supplementary Material (``Generation of trials'' and ``Supplementary results'').
\subsubsection{Questionnaires}
\label{questionnaires}
We used the Problem-Solving Inventory (PSI; \cite{heppnerDevelopmentImplicationsPersonal1982}) to assess self-reported problem-solving skills. After the problem-solving and evaluation blocks, participants also described their strategies in free-text response boxes, and the debrief questionnaire collected demographic data and feedback on the study experience.
\subsection{Measures}
\label{measures}
To make the data hierarchy explicit we distinguish four nested levels of variables. Participant-level variables are constant for each person (e.g., age, expertise). Problem-level variables take one value per problem instance (e.g., number of items and bins). Solution-pair-level variables are computed once for the two solutions taken together such as their maximum, sum, or difference --- and are therefore shared by both solutions in that trial. Solution-level variables describe a single solution within the pair (e.g., score of the left solution, format of the right). While not reported as primary measures, they are inputs to the calculation of the solution-pair-level variables presented below. All measures reported below are tagged with these level names so that their place in the data structure is unambiguous.
\subsubsection{Dependent variables}
\label{dependent-variables}
\paragraph{Choice (solution-pair-level)}
\label{sec:org2b63f01}

The outcome variable, \emph{choice}, captured participants' responses during evaluation trials using four ordered categories: \emph{definitely left}, \emph{slightly left}, \emph{slightly right}, and \emph{definitely right}. This variable was treated as an ordinal factor in all statistical analyses and coded in ascending order: \emph{definitely left} < \emph{slightly left} < \emph{slightly right} < \emph{definitely right}.
\paragraph{Reaction time (solution-pair-level)}
\label{sec:org6d4516f}
The continuous variable \emph{reaction time} (RT) is the elapsed time during the evaluation trial that a person needed to make their choice, recorded in milliseconds from stimulus presentation to participant response. The natural logarithm of reaction time was used in analyses to normalize the positively skewed distribution typical of response time data.
\paragraph{Gaze bias (solution-pair-level)}
\label{sec:orgcf6df39}

This solution-pair-level metric is quantified as the relative difference in gaze sample counts between the right and left stimuli, derived from eye-tracking data. For each trial, gaze samples are assigned to either the left (\(L\)) or the right (\(R\)) solution. For statistical analysis, these counts are modeled using a binomial generalized linear mixed model (GLMM) with a logit link function on the vector \((R, L)\). For descriptive reporting, a continuous bias value,

\begin{equation}\label{eqn-gaze-bias}
b = \frac{R - L}{R + L},
\end{equation}

\noindent is computed, ranging from -1 to 1. Trials with no valid gaze samples (\(R+L=0\)) are excluded from the analysis.
\subsubsection{Complexity models}
\label{complexity-models}
We operationalize complexity using three distinct solution-level metrics: heuristic-related complexity (HC), compositional complexity (CC), and visual-order complexity (VC) (see Figure \ref{fig:complexity}). Below, we detail the derivation of each complexity measure for a single solution. To serve as solution-pair-level predictors in our statistical models, we then compute differences between the paired solutions presented in each evaluation trial: signed differences (right minus left) for choice and gaze, and absolute differences for reaction time.
\paragraph{Heuristic-related complexity (HC, solution-level)}
\label{sec:org6bd1605}

To compute HC for a given solution, we first construct a greedy reference solution using a \emph{Largest Bin First, Largest Item First} (LBF-LIF) strategy \autocite{johnsonWorstcasePerformanceBounds1974,coffmanApproximationAlgorithmsBin1997}. This involves descendingly ordering bins by capacity and items by size. We then iterate through the sorted bins; for each bin, we greedily fill it by placing the largest available unassigned items that fit, until no more items can be placed in that bin. Ties (equal bin capacities or item sizes) are broken by preserving the original input order of bins and items. We then represent both the given and the greedy solutions as bipartite graphs (bins/items as nodes; assignments as edges) and compute their graph edit distance with unit costs for edge insertion and deletion. HC is the resulting distance, with larger values indicating deviation from the greedy reference. For statistical analyses, its signed right–left difference is denoted \(\Delta\text{HC}\) and its absolute difference \(|\Delta\text{HC}|\).
\paragraph{Diagonal dissimilarity (DD, solution-level; control covariate)}
\label{sec:orgb27b055}

Since heuristic solutions to ordered problem instances (bins and items sorted in descending order by size) often resemble a diagonal line in the assignment matrix (from the top-left to the bottom-right; see Figure \ref{fig:complexity}), we included the graph edit distance to an approximated diagonal (see ``Construction of the approximated diagonal assignment matrix'' in the Supplementary Material) as a control covariate. DD is always computed on the assignment matrix as displayed: if bins or items are visually permuted, the permuted display is compared to the diagonal reference. DD therefore captures how diagonal-like the viewed layout is. By contrast, HC is defined relative to a greedy reference that internally orders bins and items by size before assignment and is thus invariant to visual permutations. DD and HC together allow us to distinguish a preference for diagonal visual layouts (DD) from a preference for heuristic-aligned structure (HC). For statistical analyses, its signed right–left difference is denoted \(\Delta\text{DD}\) and its absolute difference \(|\Delta\text{DD}|\).
\paragraph{Compositional complexity (CC, solution-level)}
\label{sec:org3cae6e1}

This metric assesses complexity based on the composition of items in each bin. Each bin is treated as the outcome of a generative model, and we quantify how surprising that outcome is under the model. In this context, greater surprisal reflects higher complexity, as it indicates deviations from the expected patterns dictated by the model. Conversely, a low level of surprisal signifies simplicity, suggesting that the bin conforms closely to these preferred patterns. A bin is characterized by three properties: (a) the number of items, \(N\); (b) the vector of relative item sizes, \(C\), which sums to one when the bin is nonempty; and (c) the unused capacity fraction, \(E\). Assuming conditional independence, the joint density factorizes as

\begin{equation}\label{eqn-joint-probability}
p(N,\ C,\ E)\  = \ p(N)\  \cdot \ p(C\ |\ N)\  \cdot \ p(E\ |\ N).
\end{equation}
\begin{enumerate}
\item Number of Items
\label{sec:org2d0f202}

\(N\) follows a geometric law starting at zero, \(N \sim \mathrm{Geom}(p)\). The distribution assigns higher probability to small item counts; thus, bins that hold many items contribute more to the surprise score.
\item Composition of Item Sizes
\label{sec:org850d70d}

For \(N > 1\), the vector \(C\) follows a symmetric Dirichlet distribution with concentration \(\alpha\), \(C \sim \mathrm{Dir}(\alpha)\). For empty bins (\(N = 0\)) and single-item bins (\(N = 1\)), this term is absent. When \(\alpha > 1\), the model prefers evenly split item sizes; when \(\alpha < 1\), it favors one dominating item and several very small ones. An optional correction removes the baseline probability of the perfectly even split, ensuring that surprise, and thus complexity, reflects deviations from the preferred pattern rather than the size of the simplex.
\item Empty Space
\label{sec:orgd8b1a10}

The unused fraction \(E\) follows a two-component mixture distribution placing equal probability mass near 0 and 1. Consequently, bins that are almost full or almost empty are regarded as simple, whereas bins that are half-filled are deemed more surprising and therefore complex. The components can be one of three different forms. For example, in one variant we use a truncated normal mixture,

\begin{equation}\label{eqn-empty-space-norm}
E \sim \tfrac{1}{2}\,\mathrm{Norm}_{(0,1)}(0,\sigma) + \tfrac{1}{2}\,\mathrm{Norm}_{(0,1)}(1,\sigma),
\end{equation}

\noindent and analogous mixtures for the truncated Laplace and continuous Bernoulli options \autocite{loaiza-ganemContinuousBernoulliFixing2019}. A common scale parameter \(\sigma\) controls how sharply the mass concentrates around the extremes (selection of these parameters is addressed below). For a single bin, the negative log-probability

\begin{equation}\label{eqn-complexity-single-bin}
L\  = -\left[\ln p(N)\  + \ \ln p(C\ |\ N)\  + \ \ln p(E\ |\ N)\right]
\end{equation}

\noindent quantifies its surprise and therefore complexity --- in nats. With appropriate parameter settings, this formulation allows us to assign relatively low surprise to solutions consisting of few-item bins that are either nearly empty or nearly full and whose item sizes approximate symmetric compositions, while deviations from these simple patterns can be assigned higher surprise. The resulting compositional complexity of a solution is then defined as the average surprise of its bins under this model. For statistical analyses, its signed right–left difference is denoted \(\Delta\text{CC}\) and its absolute difference \(|\Delta\text{CC}|\).
\item Optimized Parameters
\label{sec:org0b0befc}

As noted above, the CC model includes several tunable parameters that influence the complexity evaluation. The empty space fraction can be described using different distributions, such as truncated normal, truncated Laplace, or continuous Bernoulli. The scale parameter controls the concentration of probability mass in these distributions. Additionally, parameter \(p\) determines how strongly the geometric law penalizes bins with many items, and parameter \(\alpha\) sets the preference for symmetry in the composition of item sizes using the Dirichlet distribution. Prior to each experiment (exploratory or confirmatory), a dedicated calibration procedure was conducted to determine these parameters (see ``Calibration of the compositional-complexity model'' in the Supplementary Material). We provide information on the parameters used in the exploratory study in the Supplementary Material (``Exploratory study''). For our confirmatory analysis, this procedure yielded the following parameters: continuous Bernoulli distribution for empty space, scale parameter \(\sigma=\) 0.426, \(p=\) 0.043 and \(\alpha=\) 0.984, with Dirichlet correction.
\end{enumerate}
\paragraph{Visual-order complexity (VC, solution-level)}
\label{sec:org6b97ccf}
This metric assesses the disorder of items and bins in a visual representation of a problem instance using an adapted version of Kendall's \(\tau\) (rank correlation). Let \(w = (w_{1},\ldots,w_{m})\) be the bin capacities and \(z = (z_{1},\ldots,z_{n})\) the item sizes. For any sequence \(a = (a_{1},\ldots,a_{k})\) \((k \ge 2)\) define

\begin{equation}\label{eqn-voc-1}
a_{i,asc} = a_{i} + i\ \varepsilon, \quad a_{i,desc} = a_{i} + (k-i + 1)\varepsilon
\end{equation}

\noindent with \(\varepsilon\  = 10^{- 5}\). The corresponding rank correlations are

\begin{equation}\label{eqn-voc-2}
\tau_{asc}(a) = \tau(a_{asc},\ (1,\ 2,\ \ldots,\ k)), \ \tau_{desc}(a)\  = \tau(a_{desc},\ (1,\ 2,\ \ldots,\ k))
\end{equation}

\noindent Disorder is quantified as

\begin{equation}\label{eqn-voc-3}
d(a) = 1\  - max(|\tau_{asc}(a)|,|\tau_{desc}(a)|).
\end{equation}

\noindent Finally, the visual-order complexity (\(VC\)) of a solution is

\begin{equation}\label{eqn-voc-4}
VC\  = \frac{m\ d(w)\  + \ n\ d(z)}{m\  + \ n},
\end{equation}

\noindent where \(m = |w|\) and \(n = |z|\).

This adaptation treats adjacent bins or items of identical size as already ordered by adding a tiny offset to break the tie (Equation \ref{eqn-voc-1}). Kendall's \(\tau\) is then computed against both an ascending and a descending reference, and the larger absolute correlation is kept (Equation \ref{eqn-voc-2}); disorder for that sequence is defined as in Equation \ref{eqn-voc-3}. Applying this procedure to the bin sequence and the item sequence and weighting the two disorder scores by their respective counts yields the VC (Equation \ref{eqn-voc-4}). This expression quantifies the overall disorder of a given visual representation of a problem instance relative to an ideally ordered state. For statistical analyses, its signed right–left difference is denoted \(\Delta\text{VC}\) and its absolute difference \(|\Delta\text{VC}|\).
\subsubsection{Further measures}
\label{further-measures}
\paragraph{Maximum disorder (MD, solution-pair-level)}
\label{sec:org47c3e2f}
This solution-pair-level metric is derived from the VC of two solutions and is defined as \(max(VC_L, VC_R)\). It captures whether at least one member of the pair is highly disordered, which may impair comparison. Because MD and VC are derived from the same underlying disorder scores, they were not entered together in later model analyses.
\paragraph{Problem difficulty (PD, problem-level)}
\label{sec:org0a2b240}
This problem-level metric was operationalized as the ratio of the sum of item sizes to the sum of all bin capacities. It ranged from 0.8 to 1.0 in the sampled problems, with higher values indicating that items collectively approach the available capacity more closely, leaving less slack and thereby increasing the demand for precise packing.
\paragraph{Heuristic optimality (HO, problem-level)}
\label{sec:orga7eb2fd}

This measure assesses the quality of a heuristic solution for a given problem instance. It is the ratio of the heuristic score to the optimal score, with 1.0 indicating that the heuristic itself attains the optimum.
\paragraph{Self-reported problem-solving skills (PSI, participant-level)}
\label{sec:org918c6bd}
This metric was calculated using the sum of participants' scores on all the items in the PSI questionnaire.
\paragraph{Problem-solving efficiency (PSE, participant-level)}
\label{sec:orge4d5e21}
This metric assesses participants' difficulty-weighted \emph{problem-solving efficiency} (PSE), integrating their solution optimality and reaction time (RT) across the seven problem-solving trials, with harder trials contributing proportionally more to the final score. For trial \(i\), unweighted efficiency was defined as

\begin{equation}\label{eqn-efficiency-1}
E_{i,j} = \frac{S_{i,j}}{O_{i}} \div RT_{i,j},
\end{equation}

\noindent where \(S_{i,j}\) is the score obtained by participant \(j\), \(O_{i}\) is the optimal score for the problem instance in trial \(i\), and \(RT_{i,j}\) is the reaction time (seconds). Trial-specific difficulty weights (\(w_{i}\)) were derived from group efficiency. Let \(n\) denote the number of problem-solving trials (here \(n = 7\)), and let \(\bar{E}_{i}\) be the mean \(E\) across participants on trial \(i\). The difficulty weights were then defined as

\begin{equation}\label{eqn-efficiency-2}
w_{i}
= \frac{1}{n - 1} \left(1 - \frac{\bar{E}_{i}}{\sum_{k=1}^{n} \bar{E}_{k}}\right),
\end{equation}

\noindent so that trials with lower mean efficiency (harder trials) received larger weights. A participant's overall PSE (\(\eta_{j}\)) was the weighted sum

\begin{equation}\label{eqn-efficiency-3}
\eta_{j}\  = \sum_{i = 1}^{n}w_{i}\ E_{i,j}.
\end{equation}

\noindent Higher values capture the ability to find solutions that are both closer to optimal and achieved more quickly, particularly on the most demanding problems.
\subsection{Data analysis}
\label{data-analysis}
\subsubsection{Data exclusion and preprocessing}
\label{data-exclusion-and-preprocessing}
To maintain data integrity, participants were excluded if they failed both catch trials, used the duplicated-solutions button in at least two non-catch trials, or did not complete the study. Individual trials were excluded if participants clicked the duplicated-solutions button, and gaze analyses additionally excluded trials with no on-stimulus gaze (\(R+L = 0\)).

To prepare for statistical analysis, all predictors were put on a comparable scale. For the three complexity measures and diagonal dissimilarity, we first computed raw right–left differences for each trial. These raw differences were then divided by their standard deviation across all trials. We did not subtract the mean of the differences, so that a value of 0 still corresponds to ``no difference between the two solutions''. Signed standardized differences were used as predictors in the choice and gaze analyses, and absolute standardized differences were used in the reaction-time analysis. The gaze outcome was not standardized. For figures and reporting, gaze is expressed as a bias, \(b=(R-L)/(R+L)\), with a range from -1 to 1.
\subsubsection{Linear mixed-effects models}
\label{linear-mixed-effects-models}
For each analysis we fitted mixed-effects models in R v4.3.2 \autocite{rcoreteamLanguageEnvironmentStatistical2023}. Predictors were entered as fixed effects, with participant random intercepts in all models and, where convergence allowed, random slopes for included complexity main effects (and DD, if present). We compared theory-motivated candidate models using the Akaike Information Criterion (AIC): if the best-fitting model was at least 2 AIC units better than every alternative (\(\Delta\text{AIC} > 2\)), it was selected; otherwise, the simplest model among those within 2 AIC units of the minimum was chosen \autocite{burnhamModelSelectionMultimodel2004}. Ordinal outcomes were analyzed with \texttt{clmm} from the \emph{ordinal} package (\cite{christensenOrdinalRegressionModels2023}; \texttt{thresholds = "symmetric"}, \texttt{link = logit}, Laplace approximation), continuous outcomes with \texttt{lmer} from lme4 (\cite{batesFittingLinearMixedeffects2015}; \texttt{REML = FALSE}), and binomial counts with \texttt{glmer} from lme4 (\texttt{family = binomial}). We report Nakagawa's marginal and conditional \(R^2\) using the \emph{performance} package \autocite{ludeckePerformancePackageAssessment2021}. 

All three analyses used right–left differences in HC, CC, VC, and DD as focal predictors and drew on the same set of potential moderators or covariates: PD, PSI, PSE, HO, and MD. Across all models, MD was never entered together with VC because both are derived from the same underlying disorder scores.

The three analyses differed in outcome, predictor form, and interaction policy. For \emph{choice}, the ordinal outcome (four levels) was predicted from signed standardized right–left differences, with two-way interactions between each complexity predictor and each moderator (no interactions among main effects or with DD). For \emph{reaction time}, the continuous outcome (log RT) was modeled using absolute between-solution differences, with the covariates included only as additional covariates and no interactions. For \emph{gaze bias}, binomial GLMMs were fitted on the counts of gaze samples on the right and left solutions, modeling the probability of gazing at the right solution given the total gaze samples on both sides; fixed effects were signed right–left differences with the same set of moderators and the same interaction policy as the choice models. Full candidate-model details are reported in the Supplementary Material (``Model fitting and comparison'').
\subsubsection{Coherence ceiling estimation}
\label{coherence-ceiling-estimation}
We inspected the three coherence trials (see \hyperref[evaluation-trials]{Evaluation Trials}) and checked whether each participant's set of pair-wise ratings was transitive. The resulting proportion of participants meeting this criterion (\(p_{coh}\)) constituted an empirical ceiling on the variance that could be attributed to stimulus properties. We therefore compared \(p_{coh}\) to the marginal Nakagawa \(R^2\) of the GLMM predicting ordinal choices from solution complexity, because the marginal \(R^2\) isolates variance explained by the fixed effect (complexity) alone, whereas the conditional \(R^2\) would also include variance due to random participant factors and would thus exceed what is theoretically explainable \autocite{nakagawaGeneralSimpleMethod2013}. This analysis was conducted with the R packages \emph{ordinal} \autocite{christensenOrdinalRegressionModels2023} and \emph{performance} \autocite{ludeckePerformancePackageAssessment2021}.
\subsection{Preregistration}
\label{sec:orgc28bbef}
We preregistered hypotheses, primary outcomes, predictors, sample size, exclusion criteria, and the analysis plan for the confirmatory study at \url{https://osf.io/d2aq7} prior to data collection. The exploratory study preceded this registration and was used to refine metrics and stimuli. Any deviations from the preregistered plan are listed below.
\subsubsection{Deviations from preregistration}
\label{sec:orgf21df89}
We implemented two deviations, described below. First, for the reaction-time (RT) analyses, we added self-reported problem-solving skills (PSI; z-scored; \autocite{heppnerDevelopmentImplicationsPersonal1982}) as a potential covariate. This corrected an oversight: PSI (self-report) and PSE (behavioral performance; preregistered potential covariate) capture complementary constructs that can both influence RT. The change affected only RT models; choice and gaze analyses remained as preregistered. In the AIC-based model selection, the final RT model did not include PSI, and including PSI as a candidate ultimately did not change the pattern of significant effects or the conclusions. Second, we modified the standardization procedure for the focal predictors (HC, CC, VC, DD). As preregistered, we planned to z-standardize the pooled left and right values and then compute the difference between these z-scores (right – left). In the final analyses, we instead computed the raw right–left differences and standardized these difference scores by dividing by their empirical standard deviation, without subtracting the mean so that 0 remained an interpretable reference point (no difference between options). This change only rescaled the predictors (and thus the regression coefficients) and did not affect standard errors, test statistics, or \emph{p}-values; the pattern of significant effects and the conclusions remained unchanged.
\section{Results}
\label{results}
\subsection{Preference for simpler solutions}
\label{preference-for-simpler-solutions}
\begin{figure*}[htbp]
  \centering \includegraphics[width=1.0\textwidth]{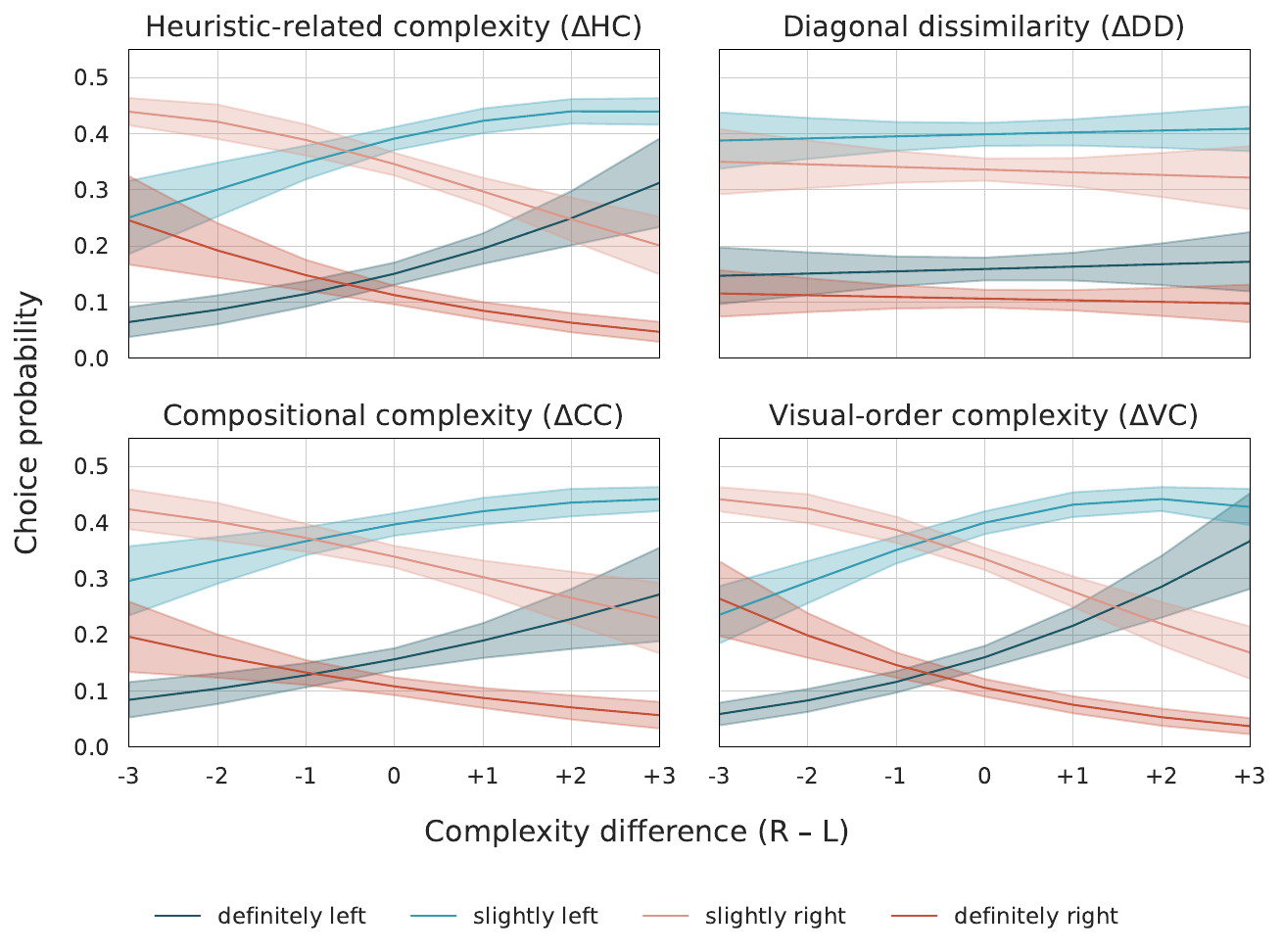}
  \caption{\label{fig:choice-probabilities-confirmatory} Predicted choice probabilities as a function of complexity difference: Panels correspond to the three complexity metrics (HC = heuristic-related complexity, CC = compositional complexity, VC = visual-order complexity) and the covariate diagonal dissimilarity (DD). Complexity differences are standardized (0 = no difference; 1 = one standard deviation). Colored lines give the model-predicted probability of the four behavioral responses (`definitely left', `slightly left', `slightly right', `definitely right'); shaded ribbons denote 95\% confidence intervals.}\end{figure*}

\begin{figure*}[htbp]
  \centering \includegraphics[height=0.8\textheight]{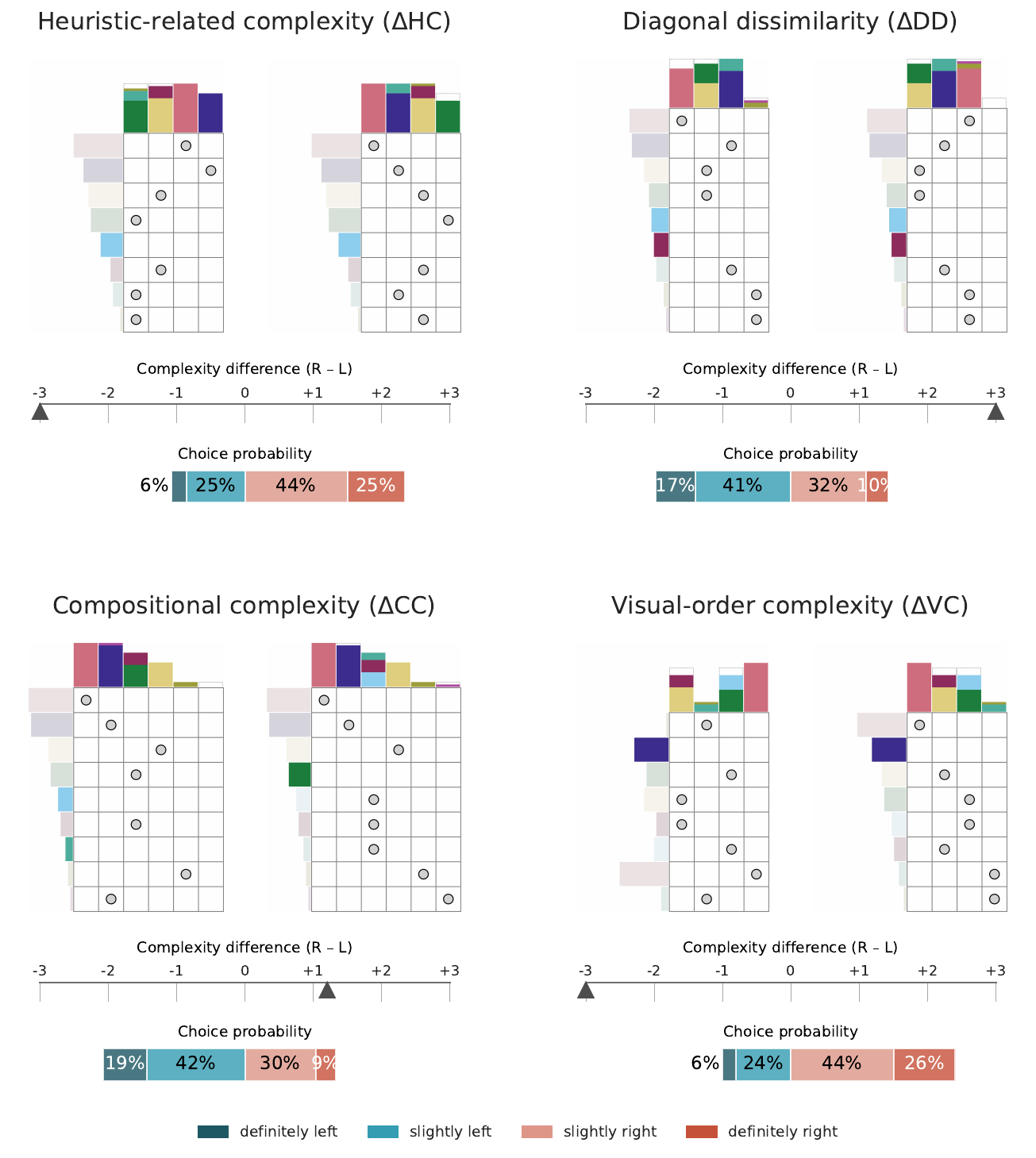}
  \caption{\label{fig:choice-examples-confirmatory}Example stimulus pairs used in the experiment across the three complexity metrics (HC, CC, VC) and the covariate diagonal dissimilarity (DD). Each panel shows the left–right solutions and their complexity difference (R–L; negative = left more complex) signified by the triangle marker. Complexity differences are standardized (0 = no difference; 1 = one standard deviation). The horizontal bar below displays model-predicted choice probabilities (\%) for definitely/slightly choosing left or right (corresponding to Figure \ref{fig:choice-probabilities-confirmatory}).}\end{figure*}

We tested standardized right–left differences in heuristic-related complexity (HC), compositional complexity (CC), visual-order complexity (VC), and diagonal dissimilarity (DD) using ordinal mixed-effects models with Akaike Information Criterion (AIC)-based selection (Table \ref{tab:choice-confirmatory}). As hypothesized, participants preferred the simpler option: all three complexity differences (HC, CC, VC) had negative coefficients (supporting H1), whereas DD did not reliably predict choice. A one-standard-deviation increase in the right–left difference reduced the odds of choosing the more complex option by 27\% for HC (OR = 0.73, 95\% confidence interval (CI) [0.64, 0.83]), 21\% for CC (OR = 0.79, 95\% CI [0.70, 0.90]), and 31\% for VC (OR = 0.69, 95\% CI [0.62, 0.77]). In practical terms, even modest one-SD differences in these metrics shifted predicted response probabilities away from the more complex option (Figure \ref{fig:choice-probabilities-confirmatory}), with the largest shifts associated with VC and HC. Because HC, CC, and VC were entered simultaneously, these effects reflect partial independence net of the other measured complexity dimensions. To illustrate these effects in terms of predicted probabilities: when all complexity differences equal zero, the model-predicted probability of choosing the right-side solution (slightly or definitely right) is approximately 45\%. A one-SD increase in the right–left complexity difference reduces this probability by 7.6 percentage points for HC, by 5.7 percentage points for CC, and by
8.8 percentage points for VC. Across 1,668 observations from 73 participants, model fit was 0.083 (marginal \(R^2\)) and 0.201 (conditional \(R^2\)). The marginal \(R^2\) remained well below the empirical coherence ceiling of 0.877 (64 participants with transitive and 9 participants with intransitive judgments), indicating that a substantial share of choice variance remains potentially attributable to systematic factors rather than mere decision noise. Figure \ref{fig:choice-probabilities-confirmatory} shows predicted probabilities shifting toward the less complex option with increasing difference, and Figure \ref{fig:choice-examples-confirmatory} illustrates representative stimulus pairs.

Across all evaluation trials, choice proportions were: definitely left 18.4\%, slightly left 38.0\%, slightly right 29.5\%, and definitely right 14.1\%. Pairwise correlations among the focal predictors were modest (\(\max |r| =\) 0.38), indicating limited collinearity (see ``Supplementary results'' in the Supplementary Material).

These confirmatory findings generally align with the exploratory analysis. However, one notable deviation was that the main effect of DD was not statistically significant in the confirmatory sample. Cross-study summary plots are shown in the Supplementary Material (``Cross-study summaries'').

\begin{table*}
\centering
\caption{Ordinal mixed-effects model predicting choice: Choice $\sim$ $\Delta$HC + $\Delta$CC + $\Delta$VC + $\Delta$DD + (1 + $\Delta$HC + $\Delta$CC + $\Delta$VC + $\Delta$DD | Subject). }
\label{tab:choice-confirmatory}
\begin{threeparttable}
\fontsize{11}{13}\selectfont
\begin{tabular}{lrrrrrr}
\toprule

Term & Estimate & SE & $z$ & $p$ & CI low & CI high \\
\midrule
\addlinespace[0.3em]
\multicolumn{7}{l}{\textbf{Fixed effects}}\\
\hspace{1em} Central Threshold & 0.136 & 0.061 & 2.239 & 0.025 & 0.017 & 0.255 \\
\hspace{1em} Threshold Spacing & 1.898 & 0.054 & 35.063 & $<\,0.001$ & 1.792 & 2.004 \\
\hspace{1em} $\Delta$HC & -0.314 & 0.064 & -4.919 & $<\,0.001$ & -0.439 & -0.189 \\
\hspace{1em} $\Delta$CC & -0.234 & 0.066 & -3.542 & $<\,0.001$ & -0.363 & -0.104 \\
\hspace{1em} $\Delta$VC & -0.371 & 0.057 & -6.541 & $<\,0.001$ & -0.482 & -0.260 \\
\hspace{1em} $\Delta$DD & -0.031 & 0.061 & -0.516 & 0.606 & -0.150 & 0.087 \\
\addlinespace[0.3em]
\multicolumn{7}{l}{\textbf{Random effects}}\\
\hspace{1em} SD (Intercept | Subject) & 0.311 &  &  &  &  &  \\
\hspace{1em} SD ($\Delta$HC | Subject) & 0.287 &  &  &  &  &  \\
\hspace{1em} SD ($\Delta$CC | Subject) & 0.360 &  &  &  &  &  \\
\hspace{1em} SD ($\Delta$VC | Subject) & 0.250 &  &  &  &  &  \\
\hspace{1em} SD ($\Delta$DD | Subject) & 0.243 &  &  &  &  &  \\
\bottomrule
\end{tabular}
\begin{tablenotes}
\item $N(obs) =$ 1668, $N(subj) =$ 73, logLik = -2101.4. The odds ratio (OR) for a 1‑SD change in a predictor is $\mathrm{OR} = \exp(\beta)$; the corresponding percent change in odds is $100\cdot(\exp(\beta) - 1)$. 
\end{tablenotes}
\end{threeparttable}
\end{table*}

\subsection{Reaction time: faster responses with larger heuristic differences}
\label{response-time-results}
We analyzed log reaction time (RT) using a linear mixed-effects model on absolute between-solution differences in HC, CC, VC, and DD (AIC-based selection). This tested the decision-speed hypothesis that larger separations facilitate choices. As hypothesized, larger \(|\Delta\text{HC}|\) predicted faster responses (partially supporting H2), amounting to an average 4\% reduction in RT per SD in \(|\Delta\text{HC}|\). In behavioral terms, solutions that differed more strongly in heuristic alignment were easier to discriminate quickly. In contrast, \(|\Delta\text{CC}|\), \(|\Delta\text{VC}|\), and \(|\Delta\text{DD}|\) did not significantly predict responses, so H2 was supported only for HC. Higher problem-solving efficiency (PSE) was also a significant predictor (Table \ref{tab:rt-confirmatory}). Fit was modest (marginal \(R^2 =\) 0.054; conditional \(R^2 =\) 0.649) across 1,668 trials from 73 participants. The corresponding raw RT had a median of 7532 ms (25th--75th percentile = 4708--12954 ms).

These results contrast with our exploratory analysis, where larger absolute differences in all three complexity metrics --- HC, CC, and VC --- were associated with faster responses. Cross-study summary plots are shown in the Supplementary Material (``Cross-study summaries'').

\begin{table*}
\centering
\caption{Linear mixed-effects model predicting response time: RT $\sim$ |$\Delta$HC| + |$\Delta$CC| + |$\Delta$VC| + |$\Delta$DD| + PSE + (1 | Subject).}
\label{tab:rt-confirmatory}
\begin{threeparttable}
\fontsize{11}{13}\selectfont
\begin{tabular}{lrrrrrr}
\toprule

Term & Estimate & SE & $t$ & $p$ & CI low & CI high \\
\midrule
\addlinespace[0.3em]
\multicolumn{7}{l}{\textbf{Fixed effects}}\\
\hspace{1em} Intercept & 9.010 & 0.068 & 131.771 & $<\,0.001$ & 8.874 & 9.146 \\
\hspace{1em} |$\Delta$HC| & -0.042 & 0.019 & -2.214 & 0.027 & -0.079 & -0.005 \\
\hspace{1em} |$\Delta$CC| & 0.016 & 0.016 & 1.013 & 0.311 & -0.015 & 0.048 \\
\hspace{1em} |$\Delta$VC| & -0.004 & 0.013 & -0.325 & 0.745 & -0.030 & 0.022 \\
\hspace{1em} |$\Delta$DD| & -0.029 & 0.016 & -1.780 & 0.075 & -0.062 & 0.003 \\
\hspace{1em} PSE & -0.167 & 0.068 & -2.460 & 0.016 & -0.303 & -0.032 \\
\addlinespace[0.3em]
\multicolumn{7}{l}{\textbf{Random effects}}\\
\hspace{1em} SD (Intercept) & 0.535 &  &  &  &  &  \\
\hspace{1em} SD (residual) & 0.411 &  &  &  &  &  \\
\bottomrule
\end{tabular}
\begin{tablenotes}
\item Outcome: log reaction time ($\log\,\mathrm{RT}$). $N(obs) =$ 1668, $N(subj) =$ 73, logLik = -1018.1. 
\end{tablenotes}
\end{threeparttable}
\end{table*}
\subsection{No evidence for complexity effects on gaze dwell times}
\label{gaze-dwell-time-no-effects-of-complexity}
We modeled side-wise dwell with a binomial generalized linear mixed-effects model (GLMM) on the counts of gaze samples on the right (R) and left (L) solutions, using a logit link. The AIC-based comparison retained the intercept-only specification, indicating no reliable complexity effects on gaze bias (H3 was not supported). The intercept was significantly negative (b = -0.400, \(p\) = < 0.001, 95\% CI [-0.612, -0.187]), consistent with a small overall leftward gaze tendency. Fit indices were low (marginal \(R^2 =\) 0.000; conditional \(R^2 =\) 0.182). Descriptively, gaze bias had a mean of -0.062 (SD = 0.467), and trials with no usable gaze comprised 4.1\%.

These results are consistent with the exploratory analysis, which likewise retained an intercept-only model.
\section{Discussion}
\label{discussion}
In this paper, we asked which properties of packing solutions to the multiple subset sum problem (MSSP) make them easier to understand. When participants compared equally optimal solutions side by side, they consistently favored the option with lower compositional complexity (CC), visual-order complexity (VC), and heuristic-related complexity (HC), supporting H1. Reaction times showed a selective speeding of decisions when heuristic-related differences were larger (partially supporting H2), whereas aggregate webcam-based gaze did not exhibit complexity-driven dwell asymmetries (H3 not supported). Together, these findings support a feature-based account of interpretability within the packing paradigm studied here and suggest practical ways to align generated solutions with human preferences.
\subsection{Interpretable structure: alignment with human heuristics and perceptual organization}
\label{sec:org105cb43}

Our results supported H1: all three complexity differences predicted choice, indicating reliable preference for simpler solutions. These convergent effects fit a simple cognitive account. First, visual order helps the perceptual system produce short, rule-like descriptions (e.g., ``largest first''), consistent with simplicity/likelihood principles in everyday perception \autocites{feldmanSimplicityPrinciplePerception2016}[][]{vanderhelmSimplicityLikelihoodVisual2000}{chaterReconcilingSimplicityLikelihood1996}{helmholtzTreatisePhysiologicalOptics1962}. Second, heuristic alignment enables immediate rationalization of how a solution was constructed, reducing explanatory burden \autocite{gigerenzerHeuristicDecisionMaking2011}. Third, compositional simplicity reduces encoding demands: extreme bin compositions (near-empty or near-full) provide summary cues that can be registered at a glance, potentially reducing the need for further perceptual processing \autocite{swellerCognitiveLoadProblem1988}.

Notably, the robust effect of HC suggests that participants may apply familiar heuristics even when evaluating completed solutions, not only when generating them. This extends heuristic accounts of problem solving \autocites{gigerenzerHeuristicDecisionMaking2011}[][]{cormenIntroductionAlgorithms2009} to the evaluation of precomputed solutions and parallels findings from discrimination paradigms in other optimization tasks \autocite{kyritsisPerceivedOptimalityCompeting2022}. Our results therefore show that alignment with a greedy packing heuristic systematically shifts interpretability preferences among equally optimal solutions.

Larger heuristic differences (\(|\Delta\text{HC}|\)) were associated with faster evaluations, consistent with H2 for HC, and with the idea that familiar construction reduces decisional conflict \autocites{gigerenzerHeuristicDecisionMaking2011}[][]{swellerCognitiveLoadProblem1988}. In contrast, differences in bin compositions and order (\(|\Delta\text{CC}|\) and \(|\Delta\text{VC}|\)) did not reliably shorten decisions, suggesting that these features guide preference without necessarily compressing total deliberation in our low-pressure setting. While the exploratory sample showed broader RT reductions, the confirmatory pattern leaves HC as the only robust speed effect \autocite{luceResponseTimesTheir1986}.

H3 predicted that complexity differences would manifest in attentional asymmetry; however, aggregate side-wise dwell did not reliably vary with complexity under webcam-based tracking, and the results suggested only a modest left-gaze tendency. In paired presentations of equally optimal alternatives, brief or small asymmetries may be swamped by inter-trial variability.

Several alternative explanations for the observed preference effects merit consideration. Preferences may partly reflect visual fluency or aesthetic regularity rather than deeper structural understanding \autocites{feldmanSimplicityPrinciplePerception2016}[][]{vanderhelmSimplicityLikelihoodVisual2000}{chaterReconcilingSimplicityLikelihood1996}{helmholtzTreatisePhysiologicalOptics1962}. Participants may also have favored solutions that resembled the strategies they themselves would use. More generally, in this untimed comparative setting, responses may not have reflected interpretability alone. However, the robust effect of heuristic alignment, which captures correspondence with a greedy construction rather than mere display geometry, argues against a purely superficial account. The DD control further helps separate diagonal visual form from heuristic-aligned structure.

These findings also suggest several practical routes for integration into optimization systems. Interpretability can be incorporated at the solution-generation stage by breaking ties among equal optima in favor of lower HC or CC, or more generally by adding small complexity penalties in multi-objective formulations \autocite{ehrgottMulticriteriaOptimization2005}. At the presentation stage, post-hoc sorting of bins and items can reduce VC without changing the underlying solution, and a shortlist of optimal or near-optimal candidates can be reranked for interpretability before display. A further step would be personalization, in which user-specific weights over HC, CC, and VC determine which optimal solution is shown.
\subsection{Limitations}
\label{limitations}
Our study has several limitations. First, the evidence comes from a single optimization problem class, a single matrix-like display format, relatively small instances (4--6 bins, 7--9 items), and a participant sample drawn from online non-experts. The present findings therefore establish interpretability effects within this packing paradigm, but they do not yet establish that the same criteria will transfer unchanged to other combinatorial domains, larger instances, professional users of optimization systems, or alternative visualizations such as schedules or network displays.

Second, our measurements of interpretability and processing were themselves constrained. As noted above, participants could have had differing interpretations of the preference elicitation prompt, and we cannot fully separate interpretability from superficial preference in the present design. Our use of webcam-based eye tracking further introduced limited spatial precision, restricting fine-grained analyses such as scanpaths and likely reducing sensitivity to subtle attentional dynamics \autocite{papoutsakiWebgazerScalableWebcam2016}.

Third, our experimental setup involved judgments of fully computed optimal solutions without time pressure. While this controlled environment enabled clean comparisons, it differs from real-world allocation and design tasks, which often involve partial solutions, dynamic constraints, deadlines, and consequences for downstream action \autocites{leeTrustAutomationDesigning2004}[][]{dietvorstAlgorithmAversionPeople2015}.

Finally, although we designed the sampling and calibration procedures to vary our metrics systematically, HC, CC, and VC remain theoretically motivated proxies rather than exhaustive descriptors of interpretable structure. Other unmeasured properties may have covaried with them, and HC in particular was defined relative to a single greedy strategy (largest-bin, largest-item first). The applicability of these findings to alternative human-plausible heuristics \autocite{johnsonWorstcasePerformanceBounds1974,coffmanApproximationAlgorithmsBin1997,kellererKnapsackProblems2004} remains to be determined.
\subsection{Future directions}
\label{sec:org7cdaf12}

Future work should prioritize more precise measurement and designs that better reflect intended real-world use. This means going beyond preference judgments to assess how well people can use machine-generated solutions. For example, participants could be asked to complete partially finished solutions, verify constraints, detect errors, or adapt solutions under changing requirements, yielding performance-based measures such as accuracy and completion time. Richer subjective assessment could also include dedicated questionnaires on perceived interpretability, cognitive load, and satisfaction \autocites{brookeSystemUsabilityScale1996}[][]{doshi-velezRigorousScienceInterpretable2017}[][]{narayananHowHumansUnderstand2018}{afsarDesigningEmpiricalExperiments2023}. Integrating laboratory eye tracking and pupillometry would further provide more sensitive process measures of attention and cognitive effort related to HC, CC, and VC \autocites{ecksteinEyeGazeWhat2017}[][]{gollanGazeQuantifyingConscious2025}. To better approximate practical allocation and design settings, future paradigms should also incorporate features such as time pressure, dynamic constraints, partial solutions, and task consequences for downstream action \autocites{leeTrustAutomationDesigning2004}[][]{dietvorstAlgorithmAversionPeople2015}.

Generalizing the present findings is a crucial next step. This involves validating the metrics across a broader range of packing and knapsack variants, larger problem instances, alternative human-plausible heuristics for HC, and different visualization styles \autocites{kellererKnapsackProblems2004}[][]{cacchianiKnapsackProblemsOverview2022}[][]{gurskiKnapsackProblemsParameterized2019}[][]{johnsonWorstcasePerformanceBounds1974}{coffmanApproximationAlgorithmsBin1997}. Directly examining presentation strategies is also important, including stepwise derivations, interactive reveals, or alternative displays of the same solution \autocites{foxExplainablePlanning2017}[][]{chakrabortiPlanExplanationsModel2017}[][]{tverskyAnimationCanIt2002}. Personalizing generation and presentation of solutions based on user-specific preferences likewise represents a promising direction for human--algorithm collaboration \autocites{millerExplanationArtificialIntelligence2019}[][]{zerilliHowTransparencyModulates2022}.

An important theoretical and practical challenge involves quantifying interpretability--optimality trade-offs by integrating interpretability terms as secondary objectives within multi-objective optimization formulations \autocite{ehrgottMulticriteriaOptimization2005}, helping to identify when people prefer simpler, objectively worse solutions. Longer term, a unified cognitive model integrating HC, CC, and VC into a single interpretability representation could explain choices, reaction times, and attention jointly; validating such a model through out-of-sample prediction and physiological process measures would offer a more comprehensive framework \autocites{luceResponseTimesTheir1986}[][]{ecksteinEyeGazeWhat2017}[][]{francoGenericPropertiesComputational2021}[][]{francoTaskindependentMetricsComputational2022}.

Participants also provided free-text descriptions of their evaluation strategies. An exploratory reading suggested recurring mentions of order, largest-first strategies, and bin fullness, broadly consistent with the quantitative effects, but these reports were not subjected to preregistered analysis. The raw responses are available in the \href{https://doi.org/10.5281/zenodo.23964950}{Zenodo repository}; a systematic qualitative analysis remains a direction for future work.
\subsection{Conclusion}
\label{sec:org89d2772}

Within the packing paradigm studied here, human preference between equally optimal machine solutions was shaped by three quantifiable structural properties: visual order, alignment with a greedy heuristic, and compositional simplicity. These findings yield actionable design principles for interpretability-aware solution presentation and optimization. For presentation, visual-order complexity can be reduced by sorting bins and items so that perceptual disorder is lower. For optimization, interpretability can be treated as a secondary criterion, for instance by preferring solutions with lower CC and HC among equally good candidates, by breaking ties in favor of lower CC and HC, by adding small penalties for complexity in multi-objective formulations \autocite{ehrgottMulticriteriaOptimization2005}, or by screening a shortlist of optimal solutions and presenting the most interpretable one. More broadly, integrating interpretability as an explicit objective alongside traditional performance criteria may help enhance transparency, accelerate appropriate human reactions, strengthen trust, and support decisive control within human-–AI interactions for problem-solving tasks.
\section*{Data availability statement}
\label{data-availability-statement}
The data, benchmark instances, and study materials that support the findings of this study are openly available on Zenodo at \href{https://doi.org/10.5281/zenodo.23964950}{doi.org/10.5281/zenodo.23964950}. The analysis code is available at \href{https://github.com/univiemops/joptim-analysis}{github.com/univiemops/joptim-analysis} and the experiment code at \href{https://github.com/univiemops/joptim-exp}{github.com/univiemops/joptim-exp}.
\section*{CRediT authorship contribution statement}
\label{credit-authorship-contribution-statement}
DP: Conceptualization, Investigation, Methodology, Software, Formal Analysis, Data Curation, Visualization, Writing -- Original Draft, Writing -- Review \& Editing. FJ: Methodology, Writing -- Review \& Editing. DS: Writing -- Review \& Editing. FS: Resources, Supervision, Writing -- Review \& Editing. FM: Supervision, Conceptualization, Investigation, Methodology, Software, Writing -- Original Draft, Writing -- Review \& Editing.
\section*{Acknowledgments}
\label{acknowledgments}
We would like to thank Rita Hansl, Alex Karner, Kathrin Kostorz, Cindy Lor, Daniel Reiter, Annika Trapple, Nicole Wimmer, and Mengfan Zhang for their contributions during the development of the web-based experiment. We thank Hermann Kaindl for helpful discussions and feedback during the development of this research.
\section*{Competing interests}
\label{competing-interests}
The authors declare no competing interests.
\section*{Funding}
\label{funding}
This research was funded by the Austrian Research Promotion Agency (FFG), Project Nos. 471030, 887474, and 927913. D.P. was funded by the Austrian Federal Ministry for Women, Science and Research (BMFWF) via the OeAD [MPC-2025-04543]. F.M. was funded by the Austrian Science Fund (FWF) [10.55776/ESP133].
\section*{Ethical statement}
\label{ethical-statement}
The studies reported in this manuscript were approved by the Ethics Committee of the University of Vienna (approval no. 01073, dated 27 February 2024). The approval covered both the exploratory and the confirmatory study, and data collection began for each study only after the approval date (exploratory study: 28 August to 12 September 2024; confirmatory study: 10 to 20 November 2025). All participants provided electronic informed consent before participation.
\section*{Declaration of generative AI and AI-assisted technologies in the manuscript preparation process}
\label{sec:orgd18e760}
This manuscript was created entirely by the authors. Language-processing tools (e.g., LLMs) were used solely to suggest minor edits for readability and concision. The authors reviewed and finalized all text, and take full responsibility for the content.

\clearpage
\section*{References}
\label{sec:org0cb4878}
\FloatBarrier
\printbibliography[heading=none]
\end{refsection}

\clearpage

\begin{center}
  {\Large\bfseries Supplementary Material for:}\\[0.6em]
  {\Large\bfseries Unpacking Interpretability: Human-Centered Criteria for Optimal Combinatorial Solutions}\\[0.8em]
  {\normalsize Dominik Pegler$^{1,2}$, Frank Jäkel$^{2}$, David Steyrl$^{1}$, Frank Scharnowski$^{1}$, and Filip Melinscak$^{1}$}\\[0.3em]
  {\footnotesize $^{1}$Department of Cognition, Emotion, and Methods in Psychology, Faculty of Psychology, University of Vienna\\
   $^{2}$Centre for Cognitive Science \& Institute of Psychology, TU Darmstadt}
\end{center}

\setcounter{tocdepth}{2}
\tableofcontents

\begin{refsection}
\addtocontents{toc}{\protect\setcounter{tocdepth}{2}}
\setcounter{section}{0}%
\setcounter{table}{0}%
\setcounter{figure}{0}%
\setcounter{equation}{0}%
\renewcommand{\thesection}{S\arabic{section}}%
\renewcommand{\thesubsection}{S\arabic{section}.\arabic{subsection}}%
\renewcommand{\thesubsubsection}{S\arabic{section}.\arabic{subsection}.\arabic{subsubsection}}%
\renewcommand{\thetable}{S\arabic{table}}%
\renewcommand{\thefigure}{S\arabic{figure}}%
\renewcommand{\theequation}{S\arabic{equation}}%
\FloatBarrier
\section{Supplementary methods}
\label{sec:orgee1e668}
\label{sec:supplementary-methods}
\subsection{Model fitting and comparison}
\label{appendix-c-model-comparison}
\label{sec:model-comparison}
\subsubsection{General conventions}
\label{c.0-general-conventions}
\begin{itemize}
\item Software and estimation: \texttt{clmm} (\texttt{thresholds = "symmetric"}, \texttt{link = "logit"}, Laplace approximation); \texttt{lmer} with \texttt{REML = FALSE}. All predictors z-standardized across the full sample.

\item Candidate model generation: Two-way interactions only and only between complexity predictors and moderators. No interactions among main effects. maximum disorder is never entered in a model that includes visual-order complexity. Predictors may enter individually; moderators enter only with their associated main effect(s).

\item Random effects (constant within each analysis): Participant random intercepts in all models; random slopes for the included complexity main effects (and diagonal, if present) where convergence allows. Interactions are not given random slopes. If the maximal structure fails, simplify uniformly across all candidates until convergence.

\item Model selection: AIC-based selection; retain the model with AIC at least 2 units lower than all competitors (\(\Delta\text{AIC} > 2\)). If multiple within 2 AIC units, we consider them equivalent \autocite{burnhamModelSelectionMultimodel2004} and choose the simplest (fewest parameters).

\item Preprocessing: choice and gaze use signed right--left differences for complexity predictors (and diagonal dissimilarity); \emph{RT} uses absolute differences. Moderators are not differenced.
\end{itemize}
\subsubsection{Choice (ordinal outcome)}
\label{c.1-choice-ordinal-outcome}
\begin{itemize}
\item Outcome: choice (right vs left; \texttt{clmm}).

\item Main effects: \(\Delta\text{HC}\), \(\Delta\text{CC}\), \(\Delta\text{VC}\), \(\Delta\text{DD}\).
\item Moderators: PD, PSI, PSE, MD, HO.
\item Interactions: \(\Delta\text{HC}/\Delta\text{CC}/\Delta\text{VC} \times\) moderators.
\item Constraint: Do not include MD with \(\Delta\text{VC}\).
\end{itemize}
\subsubsection{Reaction time (continuous outcome)}
\label{c.2-reaction-time-continuous-outcome}
\begin{itemize}
\item Outcome: log RT (\texttt{lmer}; \texttt{REML = FALSE}).
\item Main effects: \(|\Delta\text{HC}|\), \(|\Delta\text{CC}|\), \(|\Delta\text{VC}|\), \(|\Delta\text{DD}|\).
\item Additional covariates: PD, PSI, PSE, HO, MD.
\item No moderators/interactions
\item Constraint: Do not include MD with \(|\Delta\text{VC}|\).
\end{itemize}
\subsubsection{Gaze bias (binomial outcome)}
\label{c.3-gaze-bias-binomial-outcome}
\begin{itemize}
\item Outcome: \texttt{cbind(R, L)}, binomial(logit), \texttt{glmer}.
\item Main effects: \(\Delta\text{HC}\), \(\Delta\text{CC}\), \(\Delta\text{VC}\), \(\Delta\text{DD}\).
\item Moderators: PD, PSI, PSE, MD, HO.
\item Interactions: \(\Delta\text{HC}/\Delta\text{CC}/\Delta\text{VC} \times\) moderators.
\item Constraint: Do not include MD with \(\Delta\text{VC}\).
\end{itemize}
\subsection{Stimulus and trial generation}
\label{appendix-a-stimulus-and-trial-generation}
\label{sec:stimulus-trial-generation}
\subsubsection{Generation of problem and solution instances}
\label{a.1.-generation-of-problem-and-solution-instances}
\label{sec:stimulus-generation}
Problem instances were generated via a simulation process consisting of 20,000 iterations. In each simulation iteration a bin-packing problem instance was constructed by randomly determining the number of items and bins. The number of items was sampled from a discrete uniform distribution, U(7, 9), and the number of bins was sampled from U(4, 6). Item sizes were drawn from a discrete uniform distribution ranging from 5 to 100 in steps of 5. Overall bin capacities were based on a load-capacity ratio sampled uniformly from [0.8, 1.0]. We then allocated the overall bin capacity across bins while respecting a minimal bin capacity of 10, a maximal bin capacity of 100, and steps of 10. To ensure meaningful problem instances, each instance was screened to ensure that no item size exceeded the largest bin capacity, and no bin capacity was smaller than the smallest item size.

The resulting problem instance contained a vector of item loads and a corresponding vector of bin capacities. Optimal solutions for each problem instance were computed using the OR-Tools constraint programming satisfiability solver (CP-SAT; \texttt{ortools} 9.8.3296; \cite{cpsatlp}). Since the solver returned only one optimal solution by default, we iteratively added constraints to exclude previously found solutions until we obtained up to 100 distinct optimal solutions. The process stopped when either 100 solutions were found or a new solution's objective value dropped below that of the current optimal solutions.

Subsequent to optimal solution identification, redundant solutions were removed, which involved checking for equivalence in bin compositions; for instance, two solutions were deemed equivalent if they contained bins with the same set of items, such as one bin comprising items of sizes 100, 90, 80, 40 and another comprising 100, 80, 90, 40, ensuring that all bins across the solutions match in content. Finally, problem instances with only one unique optimal solution were discarded to allow comparisons between multiple solutions to the same problem instance. Additional solution-specific metrics were computed, including HC and CC, both of which are described in the main text.

These 20,000 simulations resulted in 13,269 problem instances after applying all filters.
\subsubsection{Generation of trials}
\label{a.2.-generation-of-trials}
\label{sec:trial-generation}
The subsequent trial-generation procedure was designed to yield trials for two distinct parts of the experiment: problem-solving trials and evaluation trials. For the problem-solving trials, seven problem instances were selected using quantile sampling based on their load-capacity ratio, our metric representing problem difficulty --- out of all previously generated instances. More specifically, the seven quantiles were defined at equal intervals from 0 to 1, and the corresponding problem instance was selected for each quantile. The seven resulting problem-solving trials were then complemented with an arbitrary optimal solution --- the first one produced by the CP-SAT solver for each problem instance --- to provide a clear example to the participant. These seven trials only had to be created once and were used for all participants in the experiment.

For each participant, an individual set of 25 evaluation trials was derived from the full set of generated problem instances (and their associated solution instances). These trials were constructed by pairing solution instances within each problem instance. To ensure balanced and stratified sampling across key complexity metrics, problem instances were first categorized into difficulty levels (\emph{low}, \emph{medium}, \emph{high}) based on quantiles of the load--capacity ratio. For each metric of interest (namely, absolute difference in HC and absolute difference in CC), six extremized evaluation trials were obtained by selecting, for each difficulty level, two solution pairs from the top 10th percentile with respect to the metric of interest. Moreover, three random evaluation trials were generated by stratifying sampling procedures according to problem difficulty and problem size. Another three trials were generated by cloning the previously generated random evaluation trials and applying visual manipulations (permutations of the order of items and bins): one of the three pairs received a visual manipulation on its first solution, another pair received it on its second solution, and another pair received the visual manipulation on both solutions (referred to as ``duplicated random trials'' in the codebase). By stratifying the sampling procedures by problem size, two additional random trials were conducted. Unlike before, each pair consisted of two identical solutions, but their visual presentation differed: the first pair received a visual manipulation of the second solution, while the second pair received a visual manipulation of the first solution (referred to as ``random same trials'' in the codebase). These 20 trials were sampled individually for each participant. In addition, two catch trials and three coherence trials were integrated into the evaluation trials to assess participants' attentiveness and coherence. These five trials were the same for all participants and were only sampled once. For the former, two low-difficulty problem instances featuring identical solutions were chosen so that any preference in the experiment would indicate a lack of careful engagement with the task. To generate the coherence trials, we initially filtered the problem set to identify medium-level cases in terms of problem difficulty that offered at least three optimal solutions, as coherence trials required three distinct solutions to the same problem. To ensure diversity among solutions, we assessed each candidate problem by calculating the range of a chosen metric, compositional complexity, subtracting its minimum value from its maximum. Additionally, we determined an asymmetry score by evaluating the absolute difference between the median and the mean of the metric distribution. This score helps maintain a symmetric distribution of the metric values, ensuring balanced and representative solution variability. We retained problems that met or exceeded the 95th percentile for range values to ensure substantial variability in the metric. From this refined set, we selected the problem with the smallest asymmetry score. For the chosen problem, representative solutions with minimum, median, and maximum compositional complexity values were selected and paired to create the final set of coherence trials.

The entire evaluation trial sequence was constructed by assigning specific trial slots to coherence trials and catch trials while randomly distributing the remaining evaluation trials. In total, the experimental design yielded seven problem-solving trials and a structured set of evaluation trials comprising extremized, random (with and without visual manipulation), coherence, and catch trials. This careful stratification ensured that trials were balanced with respect to solution instance complexity and problem difficulty, thereby minimizing potential sampling biases.
\subsection{Construction of the approximated diagonal assignment matrix}
\label{appendix-d-construction-of-the-approximated-diagonal-assignment-matrix}
\label{sec:approximated-diagonal}
The approximated diagonal used to compute DD, which acts as a control for HC, is a purely theoretical assignment matrix that mimics a straight diagonal line; it is not intended as a realistic or optimal solution. Its sole purpose is to provide a simple geometric baseline against which to measure a solution's deviation from a diagonal pattern. Below is Python code to create this matrix:

\begin{lstlisting}[language=Python,firstnumber=1,numbers=left]
def create_diagonal_matrix_approximated(n_rows, n_cols):  
    # 1. Create an all-zero matrix M of size rows x cols.
    M = [[0 for _ in range(n_cols)] for _ in range(n_rows)]
    
    # 2. For each row i = 0...rows - 1
    for i in range(n_rows):
        # a. col_index <- round[i * (cols - 1) / (rows - 1)]
        col_index = round(i * (n_cols - 1) / (n_rows - 1))
        
        # b. Set M[i, col_index] <- 1
        M[i][col_index] = 1
        
    # 3. Return M.
    return M
\end{lstlisting}

\noindent The interpolation in Step 2a evenly spreads the ``1'' entries from the upper-left to the lower-right corner, resulting in a staircase-like diagonal when rows \(\ne\) cols.
\subsection{Calibration of the compositional-complexity model}
\label{appendix-e-calibration-of-the-compositional-complexity-model}
\label{sec:calibration}
\subsubsection{Calibration for exploratory experiment}
\label{calibration-for-exploratory-experiment}
The mixture model for compositional complexity (CC) was calibrated before data collection because (a) the trial-generation procedure relied on the complexity values that this model assigned to each solution instance and (b) the ensuing mixed-effects analyses required a single, fixed complexity estimate for each stimulus. Calibration used a corpus of 145 previously obtained optimal solutions. To create a principled target for calibration beyond random initialization, we computed three simple indices aligned with CC's intuitions and z-standardized them across the corpus, then performed a principal-components analysis (PCA) and retained the first component (largest variance share) as a univariate compound score serving as the empirical benchmark.

For each solution, let \(x \in \{0,1\}^{n \times m}\) be the assignment matrix (items \(\times\) bins), \(z = (z_1,\ldots,z_n)\) the item-size vector, and \(w = (w_1,\ldots,w_m)\) the bin-capacity vector; for bin \(i\), let the set of assigned items be denoted as \(J_i = \{\, j \mid x_{ij} = 1 \,\}\). 
\begin{itemize}
\item \textbf{Assignment variance (AV)}: the standard deviation across bins of the number of assigned items,
\end{itemize}

\begin{equation}\label{eqn-cc-av}
\mathrm{AV} \;=\; \operatorname{sd}_{i}\!\left( \sum_{j=1}^{n} x_{ij} \right).
\end{equation}

\begin{itemize}
\item \textbf{Average discrepancy (AD)}: for bins with at least one assigned item, mean remaining headroom after the largest assigned item,
\end{itemize}

\begin{equation}\label{eqn-cc-ad}
\mathrm{AD} \;=\; \operatorname{mean}_{i \in I^\ast}\!\left( w_i \;-\; \max_{j \in J_i} z_j \right),
\quad I^\ast \;=\; \left\{\, i \,\big|\, \sum_{j=1}^{n} x_{ij} > 0 \,\right\}.
\end{equation}

\begin{itemize}
\item \textbf{Average ratio (AR)}: for bins with at least one assigned item, define \(r_i = \operatorname{mean}_{j}(z_j x_{ij}) / w_i\) (the \(\operatorname{mean}_j\) includes zeros for unassigned items), then
\end{itemize}

\begin{equation}\label{eqn-cc-ar}
\mathrm{AR} \;=\; \frac{1}{\operatorname{mean}_{i \in I^\ast}\!\big( r_i \big)}.
\end{equation}

The CC model includes three continuous parameters (penalty on number of items in a bin \(p\), scale for the empty-space mixture distribution \(\sigma\), Dirichlet concentration \(\alpha\)) and two categorical switches (empty-space distribution: normal, Laplace, or continuous Bernoulli; optional Dirichlet correction), yielding six model variants. For each variant, we fitted \((p,\sigma,\alpha)\) with bounded quasi-Newton optimization (L-BFGS-B) under constraints \(p \in (0,1)\), \(\sigma > 0\) (for the normal and Laplace variants; for the continuous Bernoulli, \(\sigma \in (0,1)\)), \(\alpha > 0\), using starting values \(p=0.50\), \(\sigma=0.05\), \(\alpha=2.00\). The loss was the negative Pearson correlation between model-predicted complexities and the PCA compound scores, so minimizing the loss maximized correspondence. The parameter set with the highest final correlation was retained for the exploratory study (trial generation and analyses).

The search identified the truncated normal empty-space, with Dirichlet correction as optimal, with parameter estimates \(p=\) 0.977 (geometric),\(\sigma=\) 0.103 (SD), and \(\alpha=\) 1.620 (Dirichlet concentration).
\subsubsection{Calibration for confirmatory experiment}
\label{calibration-for-confirmatory-experiment}
The compositional-complexity model was re-tuned after the exploratory study because (a) the trial-generation procedure relied on the complexity values that this model assigned to each solution instance and (b) using it as a predictor in the mixed-effects analyses required a single, fixed parameter specification. Calibration relied on the right-versus-left preferences expressed in the exploratory sample (n = 73 participants, 1,664 trials). For every trial the model produced a complexity estimate for each of the two alternative solutions; their difference served as the sole predictor of the ordinal choice variable (\emph{definitely left}, \emph{slightly left}, \emph{slightly right}, \emph{definitely right}). As in the exploratory calibration, three continuous parameters were optimized; the geometric penalty on the number of items in a bin, the weight given to unassigned space, and the scaling factor on the entropy term. Two categorical switches were again crossed: the distribution assumed for empty assignments (normal, Laplace, or continuous Bernoulli) and the optional Dirichlet correction, yielding six discrete model variants. For each variant the three continuous parameters were fitted with bounded optimization using starting values of 0.50, 0.05, and 2.00, respectively. The loss function was the ordinal log-loss between the four-level observed choices and the probabilities implied by the model; minimizing this quantity maximized predictive accuracy. The search identified the continuous Bernoulli empty-space with Dirichlet correction as optimal, with parameter estimates \(p=\) 0.043, \(\sigma=\) 0.426, and \(\alpha=\) 0.984. This configuration was used to generate the final trial set and was held fixed for all subsequent analyses.
\subsection{Participant exclusions}
\label{participant-exclusions}
Table \ref{tab:participant-exclusions} reports how many participants of each study were excluded under each criterion, including the overlap between the two trial-level criteria. The criteria are those stated in the main text (``Data exclusion and preprocessing''): participants were excluded if they failed both catch trials, used the duplicated-solutions button in at least two non-catch trials, or did not complete the study. One confirmatory participant was additionally excluded because they completed the study on a mobile device, contrary to the participation requirements (desktop or laptop with webcam), and the session's data were excluded from all analyses.

\begin{table}[htbp]
\centering
\small
\caption{Participant accounting for the two studies: recruitment, exclusions per criterion with overlap, and the final samples.}\label{tab:participant-exclusions}
\begin{tabular}{@{}lrr@{}}
\toprule
Account & Exploratory & Confirmatory \\
\midrule
Recruited (approved Prolific submissions) & 117 & 90 \\
Did not complete the study (no usable session) & 3 & 2 \\
Excluded for using a mobile device & 0 & 1 \\
\midrule
Remaining after removing incomplete and mobile & 114 & 87 \\
Failed both catch trials & 14 & 1 \\
Used the duplicated-solutions button in $\geq 2$ non-catch trials & 30 & 13 \\
Overlap of the two trial-level criteria & 3 & 0 \\
Excluded (trial-level criteria, total) & 41 & 14 \\
\midrule
Final sample & 73 & 73 \\
\bottomrule
\end{tabular}
\end{table}

For the exploratory study, participants were recruited in six sequential cohort launches (3, 6, 12, 24, 36, and 36 participants), and 117 participants were approved in total; three of them did not complete the study, so 114 participants completed the study and form the accounting base in Table \ref{tab:participant-exclusions}. Among these, 14 failed both catch trials and 30 used the duplicated-solutions button in at least two non-catch trials; 3 participants met both criteria, so 41 participants were excluded and 73 remained for analysis. For the confirmatory study, 90 participants were approved, 2 of whom did not complete the study; one further participant completed the study on a mobile device and their data were excluded from all analyses, so 87 participants completed the study with a usable session. Among these, 1 failed both catch trials and 13 used the duplicated-solutions button in at least two non-catch trials, with no overlap between the criteria; 73 remained for analysis.
\FloatBarrier
\section{Supplementary results}
\label{sec:orgb035101}
\label{sec:supplementary-results}

\begin{figure}[htbp]
  \centering \includegraphics[width=\linewidth]{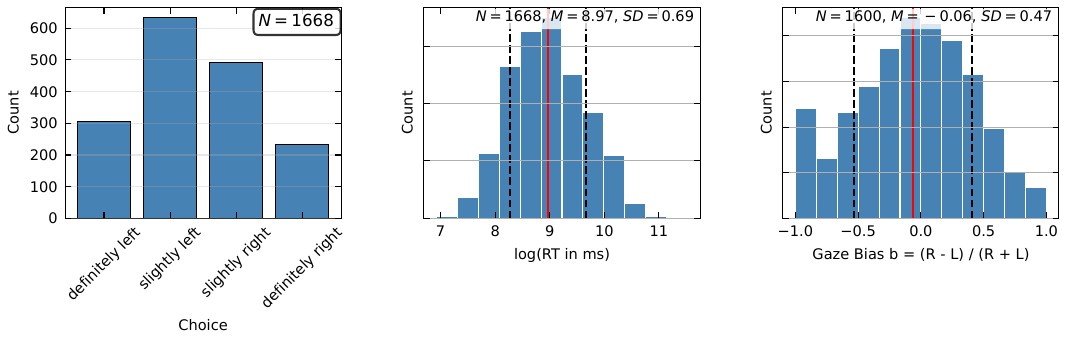}
  \caption{\label{fig:dv} Histograms show the distributions of the three dependent variables across evaluation trials: choice (four ordered categories), log reaction time ($\log\,\mathrm{RT}$), and gaze bias $b=(R-L)/(R+L)$.} 
\end{figure}

\begin{figure*}[htbp]
  \centering \includegraphics[width=\textwidth]{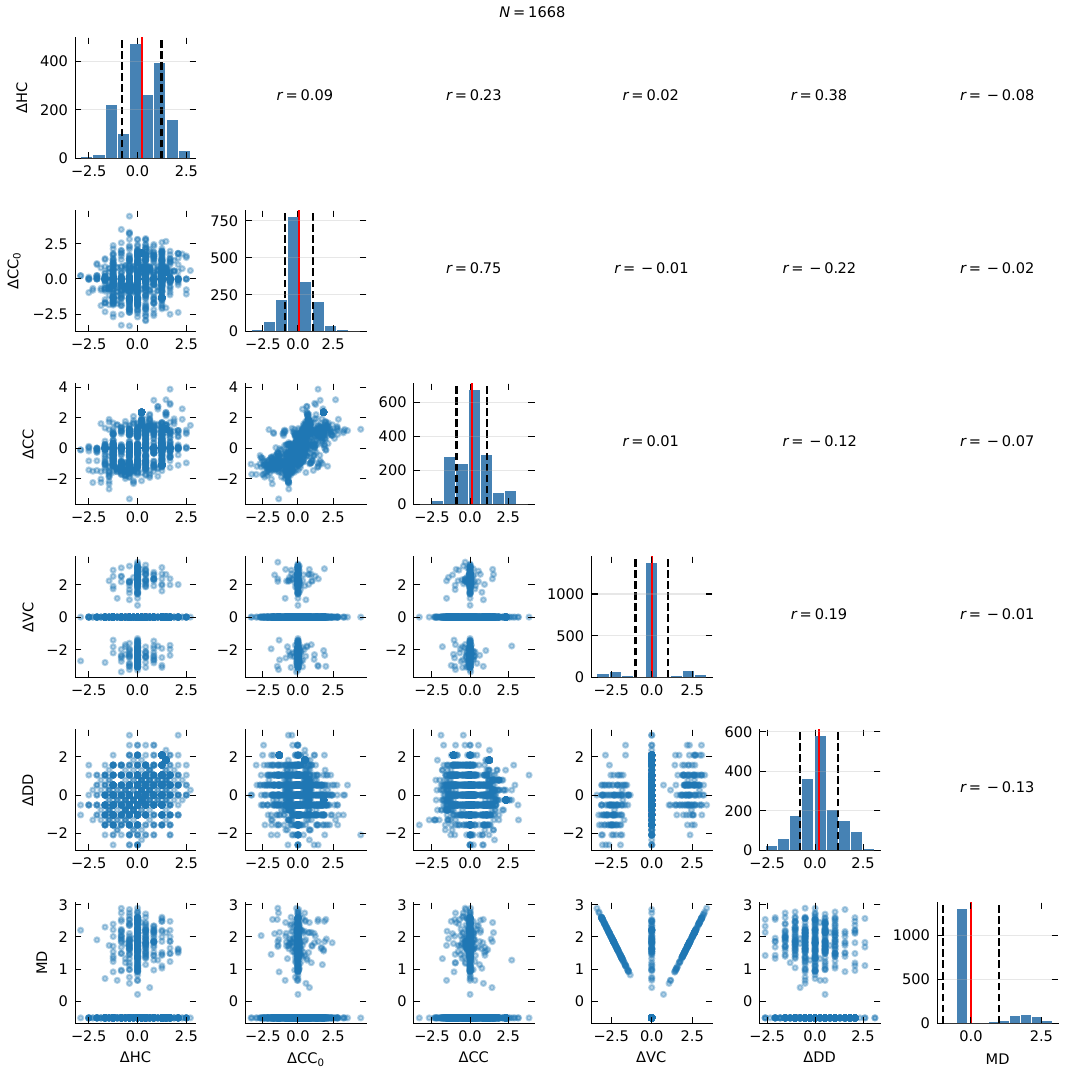}
  \caption{\label{fig:scatter-matrix-signed} Scatter matrix of solution-pair-level predictors for choice and gaze: Signed standardized differences used in the choice/gaze analyses ($\Delta$HC, $\Delta$CC, $\Delta$VC, $\Delta$DD) and Maximum Disorder (MD). CC$_{0}$ corresponds to the uncalibrated CC used in the exploratory study. Upper triangles show Pearson's $r$; diagonals show distributions (mean in red, standard deviations dashed).}
\end{figure*}

\begin{figure*}[htbp]
  \centering \includegraphics[width=\textwidth]{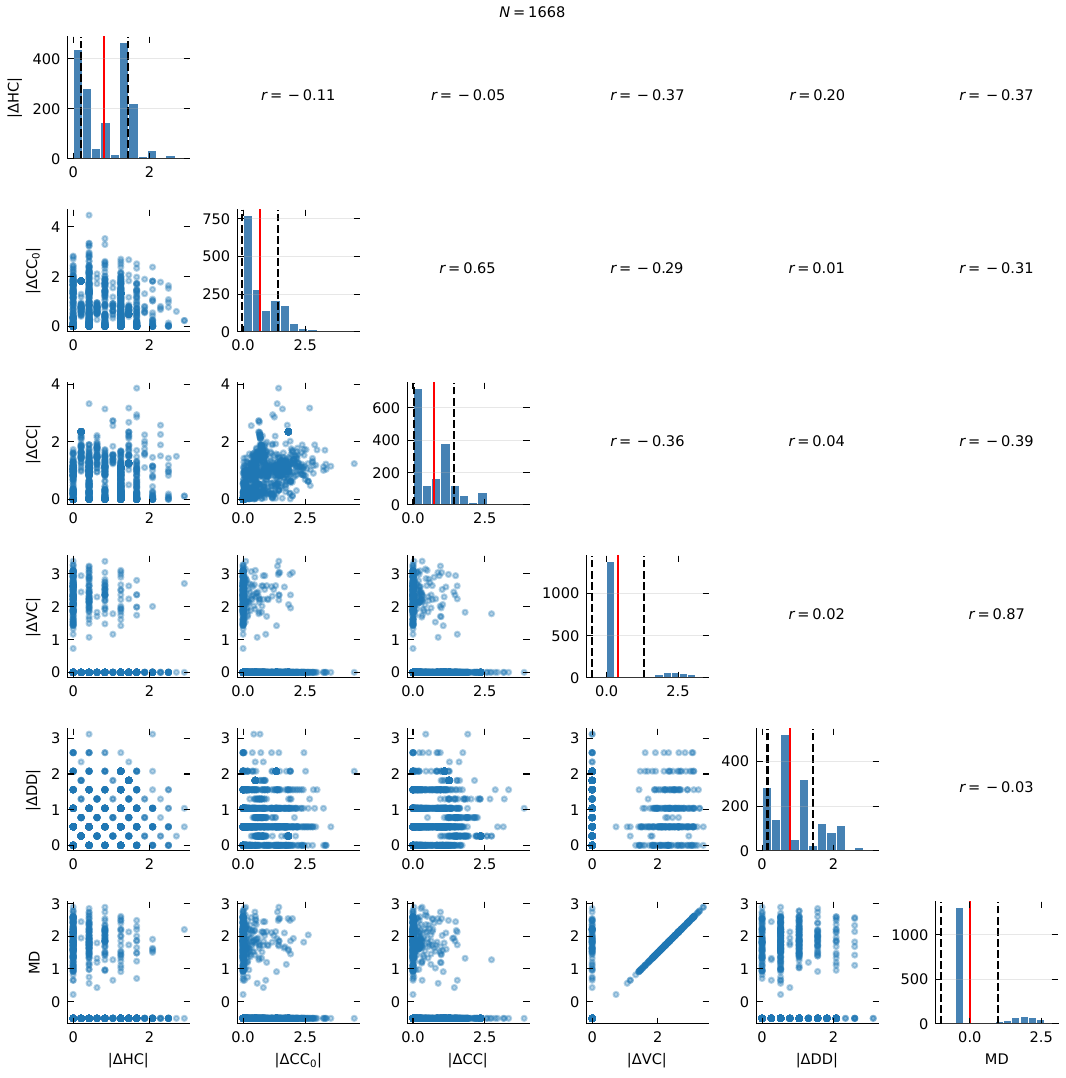}
  \caption{\label{fig:scatter-matrix-abs} Scatter matrix of solution-pair-level predictors for reaction time: Absolute standardized differences used in the RT analysis ($|\Delta|$HC, $|\Delta|$CC, $|\Delta|$VC, $|\Delta|$DD) and Maximum Disorder (MD). CC$_{0}$ corresponds to the uncalibrated CC used in the exploratory study. Upper triangles show Pearson's $r$; diagonals show distributions (mean in red, standard deviations dashed).}
\end{figure*}

\begin{figure}[htbp]
  \centering
  \includegraphics[width=0.8\linewidth]{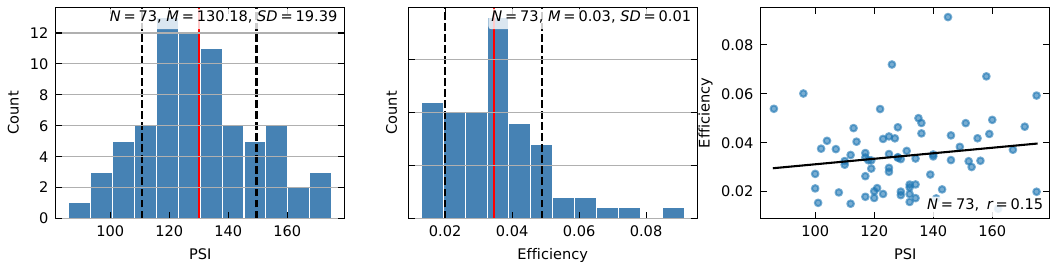}
  \caption{\label{fig:corr-participant} Distributions and correlation (Pearson) for participant-level moderators (PSI total, problem-solving efficiency). Summary statistics are reported in the text; figures document ranges and association strength used in moderation analyses.}
\end{figure}

\begin{figure}[htbp]
  \centering
  \includegraphics[width=0.8\linewidth]{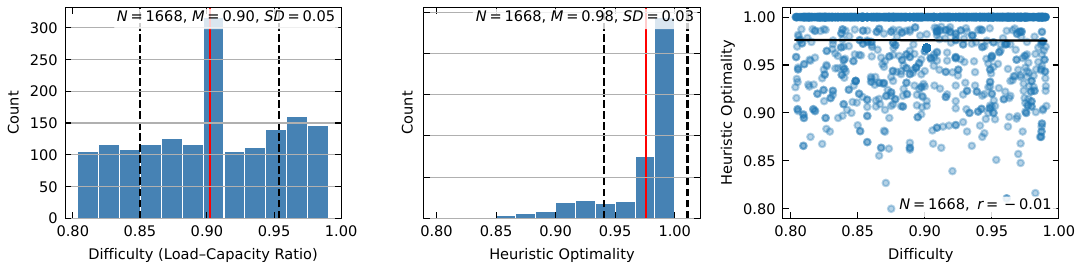}
  \caption{\label{fig:corr-problem} Distributions and correlation (Pearson) for problem-level moderators (difficulty: load–capacity ratio; heuristic optimality). Shown for transparency regarding range and potential confounding in moderation tests.}
\end{figure}
\FloatBarrier
\section{Exploratory study}
\label{appendix-b-exploratory-study}
\label{sec:exploratory-study}
\subsection{Participants}
\label{b.1.-participants}
Given the novel paradigm examined in this study, a sequential data collection approach was employed to iteratively assess outcomes and refine sample size. Initial recruitment included 3 participants to conduct a preliminary review of the study procedure and address any immediate methodological concerns. After verifying feasibility, subsequent cohorts included 6, 12, 24, 36, and 36 participants. In total, 117 participants with an approved Prolific submission were recruited across these launches; three of them did not complete the study, so 114 participants completed the study. Data collection took place from 28 August to 12 September 2024. This stepwise increase allowed for ongoing evaluation of data variability and early detection of potential effects. The final sample size was determined based on the saturation of key findings observed, with an emphasis on capturing diverse participant responses to enhance the robustness of exploratory insights. This approach was instrumental in maintaining flexibility and optimizing resource utilization throughout the study.

Of the 114 participants who completed the study, 73 participants and 1,664 evaluation trials remained after application of the exclusion criteria described in the main text; the per-criterion breakdown is reported above (``Participant exclusions''). Ages ranged from 19 to 64 years (M = 36.49, SD = 11.06), and the sample consisted of 58.90\% male and 41.10\% female. Participants took a median of 29.73 minutes to complete the experiment (25th--75th percentile = 21.73--40.80 minutes).
\subsection{Parameters for compositional complexity (CC)}
\label{sec:org5541382}

Prior to each experiment (exploratory or confirmatory), a dedicated calibration procedure was conducted to determine these parameters (see Appendix \ref{sec:calibration} for details). 
For our exploratory analysis, this procedure yielded the following parameters: truncated normal distribution for empty space, with Dirichlet correction, scale parameter \(\sigma=\) 0.103, \(p=\) 0.977 and \(\alpha=\) 1.620.
\subsection{Results}
\label{b.2.-results}
\subsubsection{Choice}
\label{choice}
\begin{figure*}[htbp]
  \centering \includegraphics[width=1.0\textwidth]{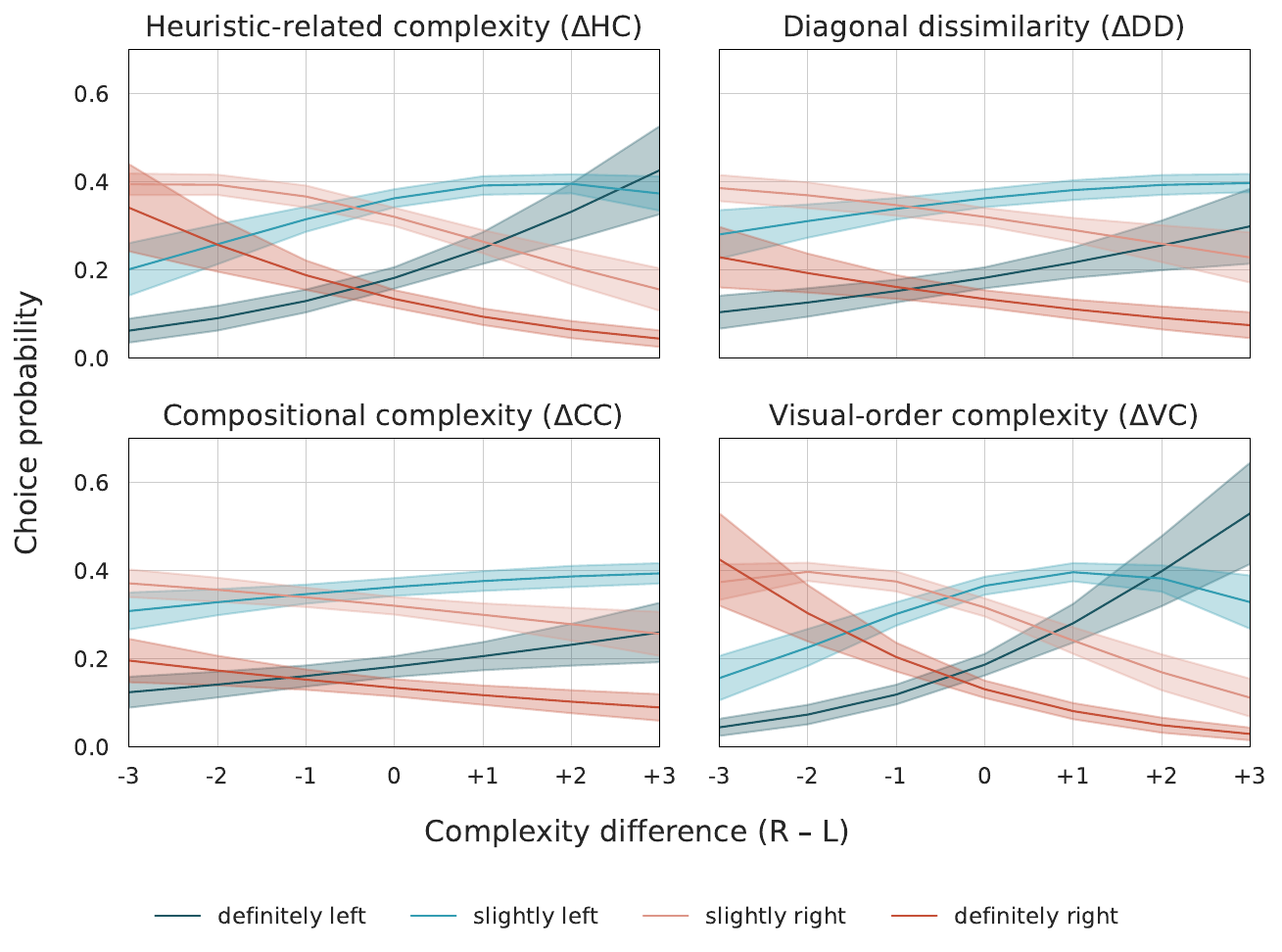}
  \caption{\label{fig:choice-probabilities-exploratory} Predicted choice probabilities as a function of complexity difference in the exploratory study: Panels correspond to the three complexity metrics (HC = heuristic-related complexity, CC = compositional complexity, VC = visual-order complexity) and the covariate diagonal dissimilarity (DD). Complexity differences are standardized (0 = no difference; 1 = one standard deviation). Colored lines give the model-predicted probability of the four behavioral responses (`definitely left', `slightly left', `slightly right', `definitely right'); shaded ribbons denote 95\% confidence intervals (CIs).}\end{figure*}

All four predictors were significant negative predictors of the ordered outcome. A one–standard-deviation increase in the difference reduced the odds of choosing the more complex solution by 33\% for HC (OR = 0.67, 95\% CI [0.59, 0.77]), 14\% for CC (OR = 0.86, 95\% CI [0.78, 0.95]), 41\% for VC (OR = 0.59, 95\% CI [0.51, 0.68]), and 19\% for DD (OR = 0.81, 95\% CI [0.71, 0.91]), with substantial between-participant variability (Table \ref{tab:choice-exploratory}). Figure \ref{fig:choice-probabilities-exploratory} illustrates how differences in stimulus complexity are reflected in participants' choice behavior. The conditional Nakagawa \(R^2\) of 0.29 indicates that the model's fixed and random effects explain about 29\% of the variance in the data, while the marginal Nakagawa \(R^2\) of 0.13 shows that the fixed effects alone account for approximately 13\%. This value is well below the coherence ceiling of 0.817, derived from the 81.69\% of participants who demonstrated transitive judgments (58 participants with transitive and 13 with intransitive judgments) in the coherence trials, indicating that there remains considerable decision variance that is not captured by the current set of structural predictors.

Across evaluation trials, choice proportions were: definitely left 22.7\%, slightly left 31.3\%, slightly right 28.7\%, definitely right 17.3\%. Pairwise correlations among the focal predictors (signed differences) were modest (max |r| = 0.33).

\begin{table*}
\centering
\caption{Ordinal mixed-effects model predicting choice in the exploratory study: Choice $\sim$ $\Delta$HC + $\Delta$CC + $\Delta$VC + $\Delta$DD + (1 + $\Delta$HC + $\Delta$CC + $\Delta$VC + $\Delta$DD | Subject). }
\label{tab:choice-exploratory}
\begin{threeparttable}
\fontsize{11}{13}\selectfont
\begin{tabular}{lrrrrrr}
\toprule

Term & Estimate & SE & $z$ & $p$ & CI low & CI high \\
\midrule
\addlinespace[0.3em]
\multicolumn{7}{l}{\textbf{Fixed effects}}\\
\hspace{1em} Central Threshold & 0.123 & 0.066 & 1.871 & 0.061 & -0.006 & 0.252 \\
\hspace{1em} Threshold Spacing & 1.681 & 0.050 & 33.533 & $<\,0.001$ & 1.583 & 1.779 \\
\hspace{1em} $\Delta$HC & -0.401 & 0.069 & -5.824 & $<\,0.001$ & -0.536 & -0.266 \\
\hspace{1em} $\Delta$CC & -0.151 & 0.050 & -2.986 & 0.003 & -0.250 & -0.052 \\
\hspace{1em} $\Delta$VC & -0.530 & 0.073 & -7.266 & $<\,0.001$ & -0.673 & -0.387 \\
\hspace{1em} $\Delta$DD & -0.216 & 0.063 & -3.444 & $<\,0.001$ & -0.339 & -0.093 \\
\addlinespace[0.3em]
\multicolumn{7}{l}{\textbf{Random effects}}\\
\hspace{1em} SD (Intercept | Subject) & 0.381 &  &  &  &  &  \\
\hspace{1em} SD ($\Delta$HC | Subject) & 0.371 &  &  &  &  &  \\
\hspace{1em} SD ($\Delta$CC | Subject) & 0.261 &  &  &  &  &  \\
\hspace{1em} SD ($\Delta$VC | Subject) & 0.440 &  &  &  &  &  \\
\hspace{1em} SD ($\Delta$DD | Subject) & 0.299 &  &  &  &  &  \\
\bottomrule
\end{tabular}
\begin{tablenotes}
\item $N(obs) =$ 1664, $N(subj) =$ 73, logLik = -2127.5. The odds ratio (OR) for a 1‑SD change in a predictor is $\mathrm{OR} = \exp(\beta)$; the corresponding percent change in odds is $100\cdot(\exp(\beta) - 1)$. 
\end{tablenotes}
\end{threeparttable}
\end{table*}

\subsubsection{Reaction time}
\label{response-time}
Larger absolute differences in HC, CC, and VC sped up choices, whereas the DD covariate had no reliable impact on reaction time (see Table \ref{tab:rt-exploratory}). The model's marginal \(R^2\) was 0.004, and the conditional \(R^2\) was 0.604, indicating modest variance explained by fixed effects with substantial additional variance captured by random participant differences. The corresponding raw RT had a median of 10812 ms (25th--75th percentile = 6320--18378 ms).

\begin{table*}
\centering
\caption{Linear mixed-effects model predicting response time in the exploratory study: RT $\sim$ |$\Delta$HC| + |$\Delta$CC| + |$\Delta$VC| + |$\Delta$DD| + (1 | Subject).}
\label{tab:rt-exploratory}
\begin{threeparttable}
\fontsize{11}{13}\selectfont
\begin{tabular}{lrrrrrr}
\toprule

Term & Estimate & SE & $t$ & $p$ & CI low & CI high \\
\midrule
\addlinespace[0.3em]
\multicolumn{7}{l}{\textbf{Fixed effects}}\\
\hspace{1em} Intercept & 9.399 & 0.075 & 124.644 & $<\,0.001$ & 9.250 & 9.549 \\
\hspace{1em} |$\Delta$HC| & -0.067 & 0.021 & -3.247 & 0.001 & -0.107 & -0.027 \\
\hspace{1em} |$\Delta$CC| & -0.041 & 0.016 & -2.552 & 0.011 & -0.072 & -0.009 \\
\hspace{1em} |$\Delta$VC| & -0.051 & 0.015 & -3.324 & $<\,0.001$ & -0.080 & -0.021 \\
\hspace{1em} |$\Delta$DD| & -0.009 & 0.019 & -0.454 & 0.650 & -0.047 & 0.029 \\
\addlinespace[0.3em]
\multicolumn{7}{l}{\textbf{Random effects}}\\
\hspace{1em} SD (Intercept) & 0.586 &  &  &  &  &  \\
\hspace{1em} SD (residual) & 0.476 &  &  &  &  &  \\
\bottomrule
\end{tabular}
\begin{tablenotes}
\item Outcome: log reaction time ($\log\,\mathrm{RT}$). $N(obs) =$ 1664, $N(subj) =$ 73, logLik = -1256.5. 
\end{tablenotes}
\end{threeparttable}
\end{table*}
\subsubsection{Gaze bias}
\label{gaze-dwell-time}
Based on the AIC-based model selection procedure, the intercept-only model was selected. The intercept was -0.385 (SE = 0.106, 95\% CI [-0.594, -0.177], \(p=\) < 0.001). Model fit indices were low (marginal \(R^2 =\) 0.000; conditional \(R^2 =\) 0.166), consistent with the intercept-only result and the absence of reliable complexity effects on side-wise dwell. Gaze bias (b) averaged -0.055 (SD = 0.457); trials with no usable gaze comprised 5.4\%. The analysis was based on 1,574 observations from 69 participants. These results indicate no evidence that between-solution right–left differences in complexity or DD affected gaze in the exploratory study. 
\subsubsection{Descriptives}
\label{sec:orgda5a3bd}

\begin{figure}[htbp]
  \centering \includegraphics[width=\linewidth]{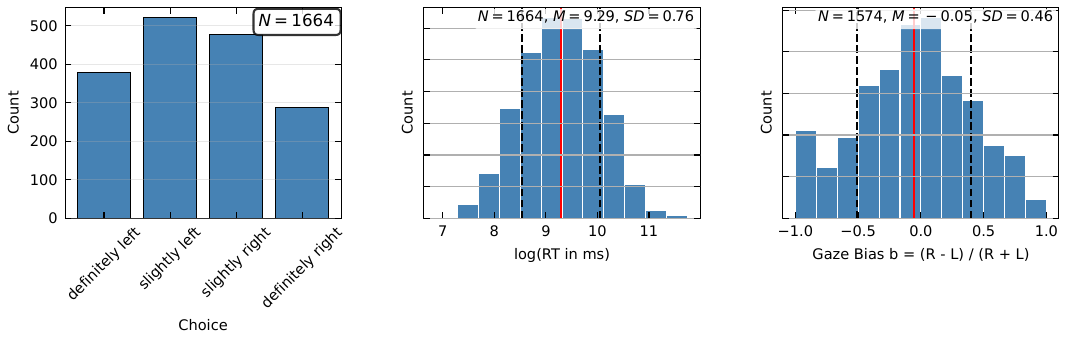}
  \caption{\label{fig:dv-e} Histograms show the distributions of the three dependent variables across evaluation trials: choice (four ordered categories), log reaction time ($\log\,\mathrm{RT}$), and gaze bias $b=(R-L)/(R+L)$ in the exploratory study.}
\end{figure}

\begin{figure*}[htbp]
  \centering \includegraphics[width=\textwidth]{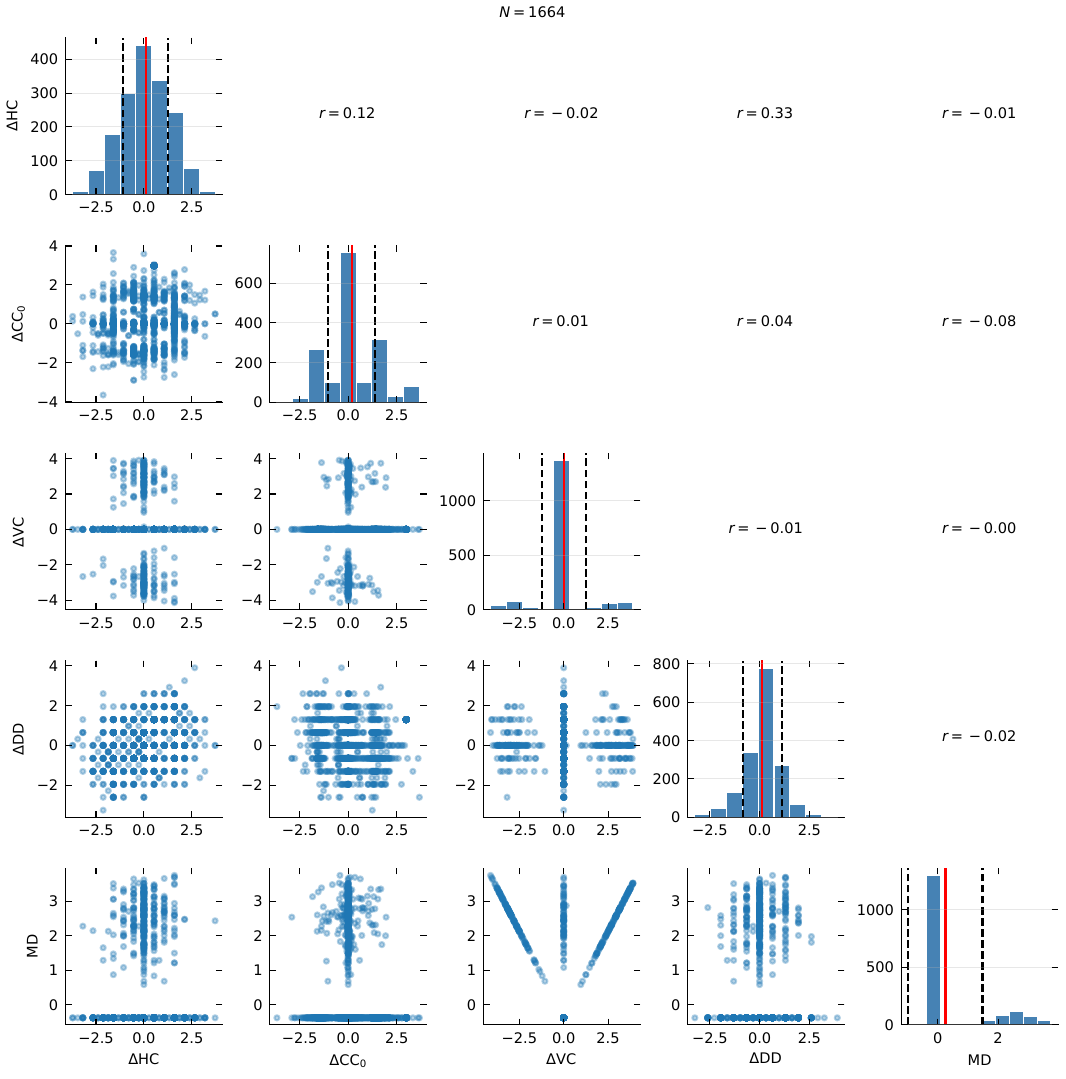}
  \caption{\label{fig:scatter-matrix-signed-e} Scatter matrix of solution-pair-level predictors for choice and gaze in the exploratory study: Signed standardized differences used in the choice/gaze analyses ($\Delta$HC, $\Delta$CC, $\Delta$VC, $\Delta$DD) and Maximum Disorder (MD). CC$_{0}$ corresponds to the uncalibrated CC used in the exploratory study. Upper triangles show Pearson's $r$; diagonals show distributions (mean in red, standard deviations dashed).}
\end{figure*}

\begin{figure*}[htbp]
  \centering \includegraphics[width=\textwidth]{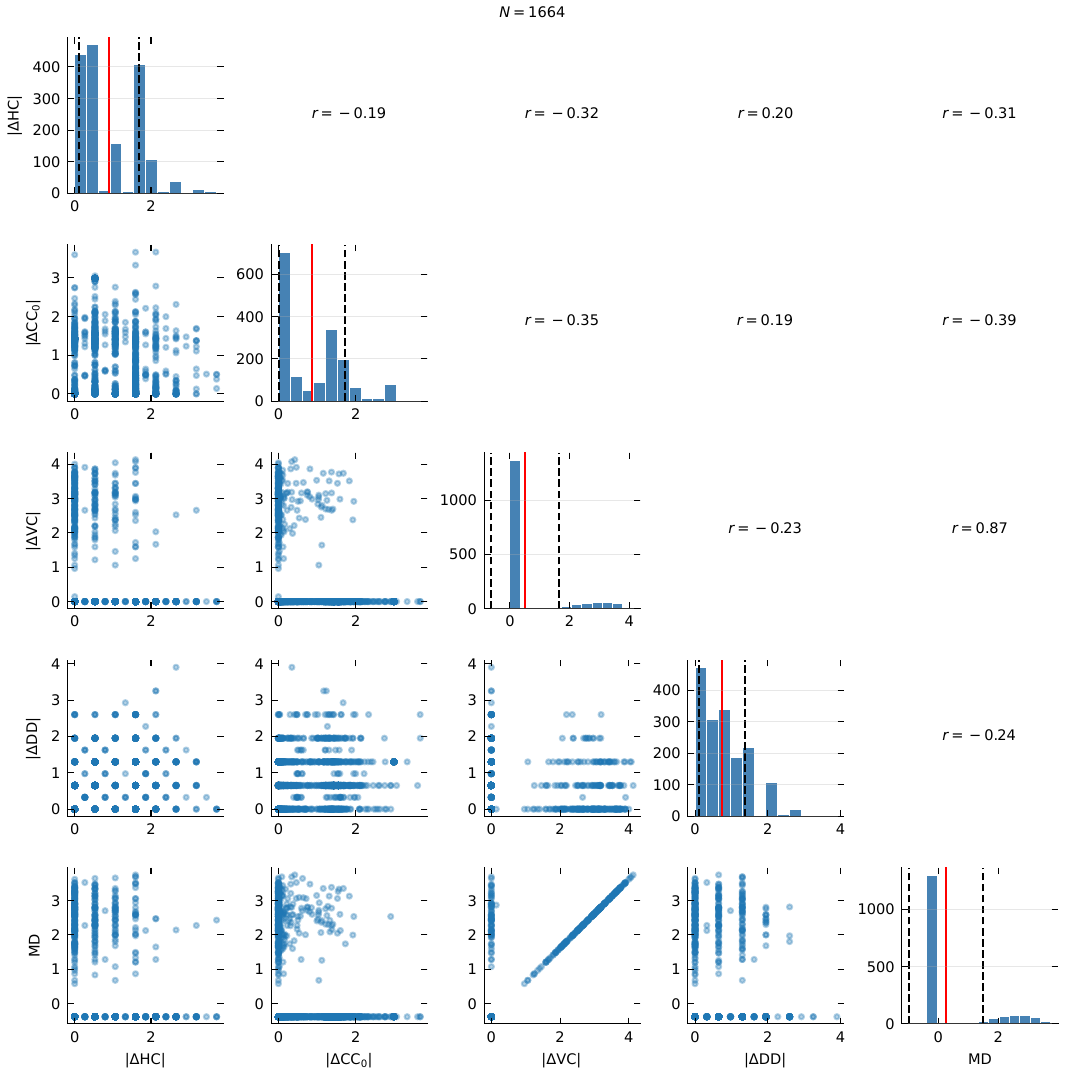}
  \caption{\label{fig:scatter-matrix-abs-e} Scatter matrix of solution-pair-level predictors for reaction time in exploratory study: Absolute standardized differences used in the RT analysis ($|\Delta|$HC, $|\Delta|$CC, $|\Delta|$VC, $|\Delta|$DD) and Maximum Disorder (MD). CC$_{0}$ corresponds to the uncalibrated CC used in the exploratory study. Upper triangles show Pearson's $r$; diagonals show distributions (mean in red, standard deviations dashed)}.
\end{figure*}

\begin{figure}[htbp]
  \centering
  \includegraphics[width=0.8\linewidth]{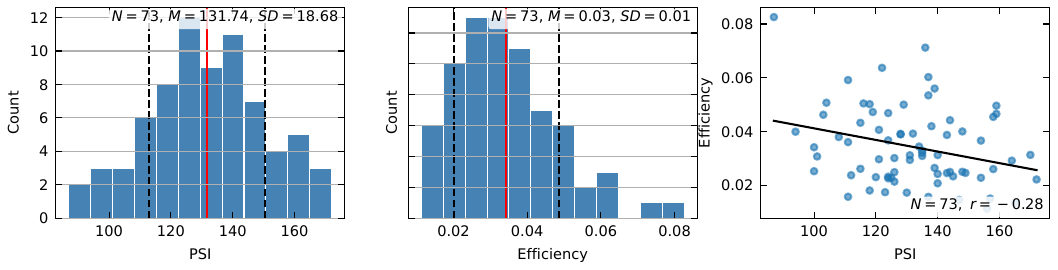}
  \caption{\label{fig:corr-participant-e}Distributions and correlation (Pearson) for participant-level moderators (PSI total, problem-solving efficiency) in the exploratory study. Summary statistics are reported in the text; figures document ranges and association strength used in moderation analyses.}
\end{figure}

\begin{figure}[htbp]
  \centering
  \includegraphics[width=0.8\linewidth]{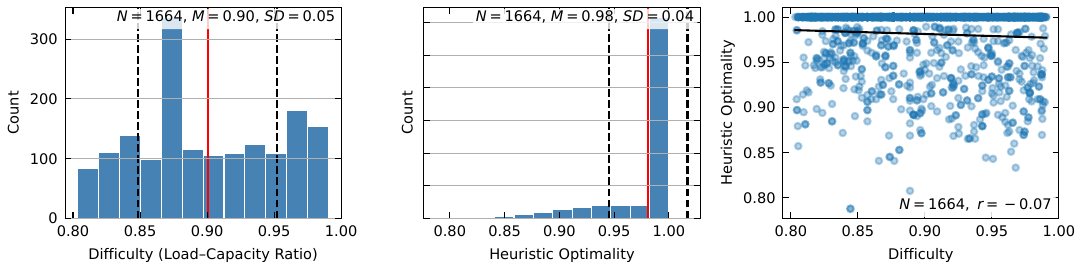}
  \caption{\label{fig:corr-problem-e} Distributions and correlation (Pearson) for problem-level moderators (difficulty: load–capacity ratio; heuristic optimality) in the exploratory study. Shown for transparency regarding range and potential confounding in moderation tests.}
\end{figure}
\FloatBarrier
\section{Cross-study summaries}
\label{sec:orga5256b3}
\label{sec:cross-study-summary}
\begin{figure}[htbp]
  \centering\includegraphics[width=0.75\linewidth]{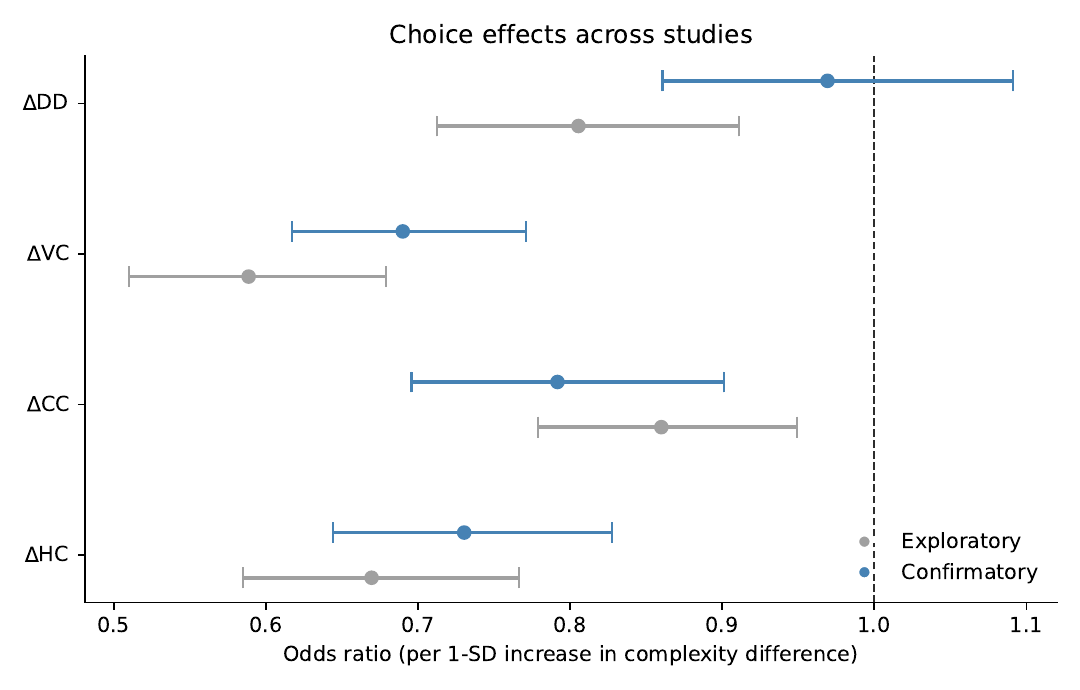}
  \caption{Cross-study odds ratios (ORs) with 95\% CIs for the effect of complexity difference on choice (HC, CC, VC, DD). Points show models fit separately to exploratory and confirmatory data; the dashed line marks OR = 1.}
\end{figure}

\begin{figure}[htbp]
  \centering\includegraphics[width=0.75\linewidth]{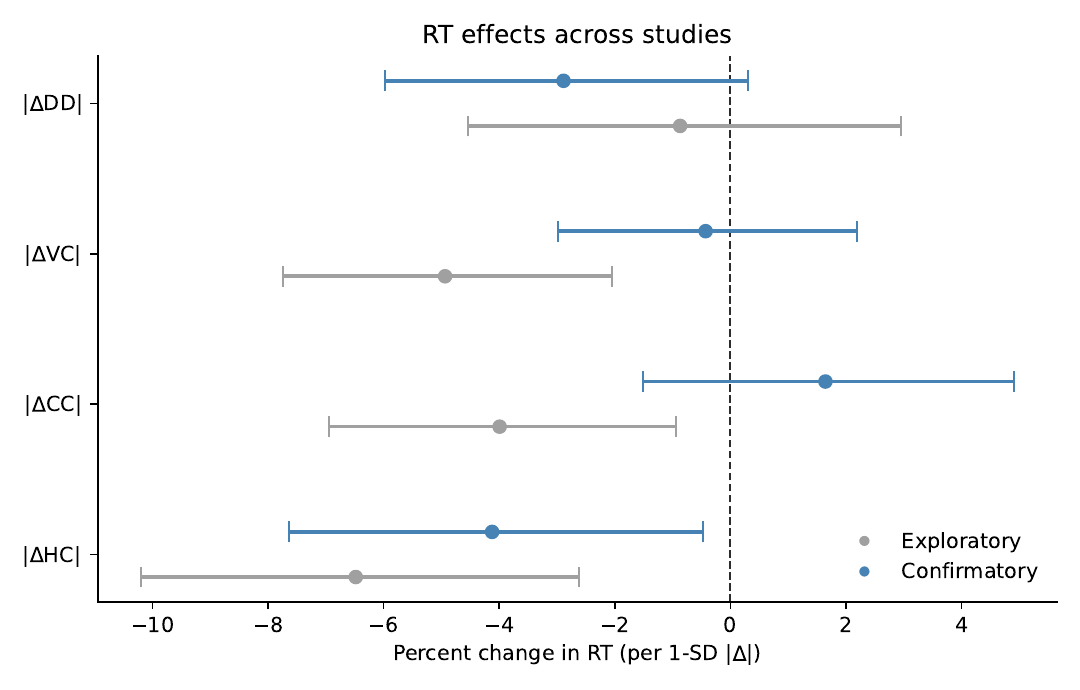}
  \caption{Percent change in reaction time per 1-SD absolute difference (|$\Delta$|) in HC, CC, VC, DD, with 95\% CIs, across exploratory and confirmatory studies. The dashed line marks 0\%.}
\end{figure}

\clearpage
\section*{References}
\label{sec:org18366c1}
\FloatBarrier
\printbibliography[heading=none]
\end{refsection}

\clearpage
\end{document}